\documentclass{aa}  

\makeatletter

\let\AA@old@journalname\aa@journalname
\let\AA@old@manuscriptname\aa@manuscriptname
\let\AA@old@AALogo\AALogo
\let\AA@old@today\today
\newlength\AA@old@fboxrule
\newlength\AA@old@fboxsep

\newcommand*\AA@disableTitleBanner{%
	\renewcommand*\aa@journalname{}
	\renewcommand*\aa@manuscriptname{}
	\renewcommand*\AALogo{}
	\def\today{}
	\setlength{\AA@old@fboxrule}{\fboxrule}
	\setlength{\AA@old@fboxsep}{\fboxsep}%
	\setlength{\fboxrule}{0pt}%
	\setlength{\fboxsep}{0pt}%
}

\newcommand*\AA@restoreTitleBanner{%
	\let\aa@journalname\AA@old@journalname
	\let\aa@manuscriptname\AA@old@manuscriptname
	\let\AALogo\AA@old@AALogo
	\let\today\AA@old@today
	\setlength{\fboxrule}{\AA@old@fboxrule}%
	\setlength{\fboxsep}{\AA@old@fboxsep}%
}

\let\AA@old@maketitle\maketitle
\renewcommand*\maketitle{%
	\AA@disableTitleBanner
	\AA@old@maketitle
	\AA@restoreTitleBanner
}

\makeatother

\usepackage[utf8]{inputenc}
\usepackage[varg]{txfonts}  
\usepackage{bm}
\usepackage{amssymb}
\usepackage{amsmath}
\usepackage{mathrsfs}
\usepackage{graphicx}
\usepackage{tabularx}
\usepackage{enumitem}
\usepackage{color}
\usepackage{subcaption}
\usepackage{hyperref}
\usepackage{booktabs}
\usepackage{multirow}
\usepackage{array}
\usepackage{upgreek}

\usepackage{fancyhdr}
\fancypagestyle{plain}{%
	\fancyhf{}                 
	\fancyfoot[C]{\thepage}    
	
}
\usepackage{appendix}
\makeatletter
\renewcommand\appendix{
	\par
	\setcounter{section}{0}%
	\setcounter{subsection}{0}%
	\setcounter{subsubsection}{0}%
	\setcounter{figure}{0}%
	\setcounter{table}{0}%
	\renewcommand{\thesection}{Appendix \Alph{section}}%
	\renewcommand{\thefigure}{\thesection.\arabic{figure}}%
	\renewcommand{\thetable}{\thesection.\arabic{table}}%
}
\makeatother

\hypersetup{
	colorlinks = true,
	linkcolor = blue,
	citecolor = blue,
	filecolor = blue,
	urlcolor = blue
}

\def\apjl{ApJL}
\def\apjs{Astrophys. J. Supp.}

\def\aap{Astron. Astrophys.}

\newcolumntype{K}[1]{>{\centering\arraybackslash}p{#1}}

\let\OLDthebibliography\thebibliography
\renewcommand\thebibliography[1]{
	\OLDthebibliography{#1}
	\setlength{\parskip}{0pt}
	\setlength{\itemsep}{0pt plus 0.3ex}
}

\allowdisplaybreaks

\title{Black hole mass function shift in proto-stellar clusters driven by gas accretion}

\author{Zacharias Roupas\inst{1,2}\thanks{Zacharias.Roupas@unimib.it}}

\institute{Dipartimento di Fisica ``G. Occhialini'', 
	Universit\'a degli Studi di Milano-Bicocca, Piazza della Scienza 3, 20126 Milano, Italy
	\and
	Istituto Nazionale di Fisica Nucleare (INFN), Sezione di Milano-Bicocca, 
	Piazza della Scienza 3, 20126 Milano, Italy
}

\date{}

\abstract{
	The James Webb Space Telescope (JWST) has observed compact, massive proto-stellar clusters of low metallicity in the Cosmic Gems arc galaxy at high redshift, which represent likely precursors to globular clusters.
	We model the mass growth of stellar black holes (BHs) during the first few million years of the life of a massive, compact, gaseous stellar cluster before stellar feedback expels the primordial gas. 
	At high redshift, in a lower-metallicity environment stellar winds get weaker, allowing for larger gas-depletion timescales in the cluster despite energetic pair-instability supernova (PISN) feedback for sufficiently compact clusters.
	Mass segregation drives the massive stellar progenitors of BHs in the center of the cluster, where gas is densest.
	We estimate the conditions for which the initial black hole mass function (BHMF), with a PISN-induced cutoff of $<55{\rm M}_\odot$, gets shifted to values within the upper BH mass gap, $\sim 60-130{\rm M}_\odot$, or higher, as observed by the LIGO-Virgo-KAGRA gravitational wave (GW) experiments.
	We find that the BHs are shifted by the end of gas depletion to values within and above the mass gap, well within the range of BH components of the recent GW-signal GW231123, depending on the total mass, star formation efficiency, metallicity, and compactness. The individual BH mass increase approximately follows a surprisingly steep power law with respect to the initial BH mass with an exponent in the range of $\approx 4-6$. This occurs in gaseous proto-stellar clusters that are sufficiently massive and compact, with typical values for the total mass of $\sim 10^6{\rm M}_\odot$ and size of $\sim 1{\rm pc}$. Our analysis suggests that proto-stellar clusters at high redshift such as Cosmic Gems arc clusters have generated through early gas accretion, BHs as heavy as $\sim 10^2-10^3{\rm M}_\odot$.
}

\begin{document}
	
	\maketitle
	
	\section{Introduction}\label{sec:intro}
	
	Recent JWST observations reveal the existence of proto-stellar clusters in the Cosmic Gems arc galaxy that are at high redshift, $z = 10.2$, younger than $50{\rm Myr}$, of low metallicity, and remarkably compact, with typical masses of $\sim 10^6{\rm M}_\odot$ and half-light radii of $\sim 1 {\rm pc}$ \citep{2024Natur.632..513A}. 
	Such proto-stellar clusters are plausibly the progenitors of globular clusters (GCs) \citep{2015MNRAS.454.1658K,2017MNRAS.469L..63R,2018MNRAS.477..480Z,2018ApJ...869..119E,2020MNRAS.491.1093V}, while they may be also relevant to the formation of nuclear star clusters (NSCs) \citep{2016MNRAS.461.3620G,2020A&ARv..28....4N}.
	These JWST-observed proto-stellar clusters share similar characteristics with super star clusters (SSCs), considered to be young GCs, observed in the past by Hubble Space Telescope (HST) at low-$z$, though they are somewhat less, but comparably, compact, have subsolar metallicities, and are observed at an age ($<5{\rm Myr}$) when they still have residual gas content \citep{1996ApJ...466L..83H, 1997ApJ...479L..27D}.
	
	Therefore, the existence of proto-stellar clusters, i.e., compact, massive stellar clusters with low or subsolar metallicity that retain residual gas for several million years is established observationally. It is possible that BHs in these clusters have enough time to accrete gas before stellar winds and SN explosions expel the primordial gas \citep{2019A&A...632L...8R}. Lower metallicity implies weaker stellar winds. Higher compactness allows more pristine gas to survive the SN explosions. The progenitors of BHs are the most massive stars, and therefore accumulate in the core of the cluster, where gas density is higher, due to mass segregation \citep{Spitzer_1987degc,2014MNRAS.441..919L}. 
	
	Identifying formation mechanisms of BHs with masses of $60{\rm M}_\odot \lesssim m_{\rm BH} \lesssim 10^3{\rm M}_\odot$ is a topical, timely, flourishing research subject in astrophysics and especially GW astronomy \citep{2017PhRvD..95l4046G,2019A&A...632L...8R,2020PhRvD.102d3002B,2020ApJ...904L..13R,Kimball_2020,2020ApJ...903L..21S,2021MNRAS.505..339M,2022MNRAS.512..884R,2023MNRAS.526..429A,2024A&A...688A.148T,2024Sci...384.1488F,2024AJ....167..191P,2025PhRvD.111f3039B,2025ApJ...988...15L,2025arXiv250808558B}.
	Gravitational wave (GW) experiments, the LIGO Scientific Collaboration, the Virgo Collaboration, and the KAGRA Collaboration \citep{2023PhRvX..13d1039A} have revealed a significant population of BHs within the theorized upper BH mass gap of $60{\rm M}_\odot \lesssim m_{\rm BH} \lesssim 130{\rm M}_\odot$, induced by the physics of a pair-instability supernova (PISN) \citep{2016A&A...594A..97B,2017MNRAS.470.4739S,2021ApJ...912L..31W}. During the completion of this work, LIGO-Virgo-KAGRA reported the GW signal GW231123 \citep{2025arXiv250708219T} involving two heavy BHs with masses of $103_{-52}^{+20} {\rm M}_\odot$ and $137_{-17}^{+22} {\rm M}_\odot$. The typical formation channel for such masses is the repeated mergers channel \citep{2002MNRAS.330..232C,2006ApJ...637..937O,2017PhRvD..95l4046G}. Here, we suggest another, not mutually exclusive, formation channel.
	
	A second reason for the significance of this BH mass range is the potential to explain the observed abundance and masses of supermassive BHs (SMBHs) at high redshifts \citep{2023ApJ...953L..29L, 2024NatAs...8..126B, 2024OJAp....7E..72R}. This issue is also closely related to the Laser Interferometer Space Antenna (LISA) astrophysics research \citep{2023LRR....26....2A}. 
	
	The major goals of this work are firstly to identify the physical conditions under which the stellar BHs assumed to be generated with a mass cutoff, $m_{\rm BH} < 55{\rm M}_\odot$, inside proto-stellar clusters accrete sufficient amount of gas to populate the upper BH mass gap before the gas gets depleted due to stellar formation feedback, and secondly to quantify this shift of the BH mass function of a BH population in a proto-stellar cluster. We find that under certain conditions it is possible that a BH grows as massive as $\sim 10^3{\rm M}_\odot$.
	We therefore provide not only a new physically motivated formation channel of mass-gap-BHs, but also a pathway from stellar-mass to intermediate-mass-BH (IMBH) seeds.
	
	In the next section we calculate gas depletion timescales. In section \ref{sec:model} we present our physical model. In section \ref{sec:results} we discuss our results. We conclude in the final section.

	\section{Gas depletion timescale}\label{sec:tau_dep}
	
	Gas expulsion in gaseous stellar clusters is driven by the energy and momentum from stellar winds and supernovae injected to the ambient gas. Additionally, the stellar lifetimes determine not only the moment of time where SNe occur, and therefore the time distribution of their energy injection, but also the birth time of BHs from their stellar progenitors. We analytically modeled the stellar winds and SN energy injection, and the stellar lifetimes, with respect to stellar mass and metallicity, according to observational and numerical data. We estimated the depletion timescale as the moment of time when the accumulated injected energy equals the initial binding energy of the gas, counting from the time when all stars are born.
	
	\subsection{Winds}\label{sec:tau_winds}
	
	The stellar winds of solar-type intermediate-mass stars of $m < 8{\rm M}_\odot$ contribute negligible energies within the depletion timescales of $\sim 0.1- 10{\rm Myr}$ (e.g., \cite{1999ApJS..123....3L}) with respect to the more massive stars of $m > 8{\rm M}_\odot$. Therefore, we considered only the latter cases.
	
	We modeled the injected mechanical luminosity of stellar winds using a piecewise power fit \citep{2002ApJ...577..389K,2008A&ARv..16..209P},
	\begin{equation}
		\frac{dE}{dt} = \left(\frac{Z}{Z_\odot}\right)^{f_Z} \cdot L_{\rm ref} \cdot \left(\frac{m}{m_{\rm ref}}\right)^\alpha ,
	\end{equation}
	where  $L_{\rm ref}$ is the reference mechanical luminosity at solar metallicity for the reference mass scale, $m_{\rm ref}$, and $L_{\rm ref}$, $\alpha$ , and $m_{\rm ref}$ take different values in three distinct mass segments,
	\begin{equation}\label{eq:wind_fit}
		(L_{\rm ref}, \,\alpha,\, m_{\rm ref} ) = \left\lbrace
		\begin{array}{llll}
			(\, 7\cdot 10^{36} & ,\, 2.42 & ,\, 20  \,)
			& ,8 \leq m < 60 \\
			(\, 10^{38}  & ,\, 1.32 & ,\, 60 \,) 
			& ,60 \leq m < 120 \\
			(\, 2.5\cdot 10^{38}  &  ,\, 1.0 & ,\, 120 \,)
			& ,m \geq 120 
		\end{array}
		\right.
		.\end{equation}
	We used these values based on existing data \citep{2001A&A...369..574V,2004A&A...415..349R,2006A&A...457.1015H, 2007ARA&A..45..177C,2011A&A...530L..14B,2017RSPTA_37560269V, 2017A&A...603A.118R, 2020MNRAS.499..873S} and to ensure continuity in the breakpoints allowing for efficient numerical operations.  
	In more detail, we assumed that
	$L_{\rm ref} = 7\cdot 10^{36}{\rm erg}/{\rm s}$ for $m_{\rm ref} = 20{\rm M}_\odot$ \citep{2004A&A...415..349R},  
	$L_{\rm ref} = 10^{38}{\rm erg}/{\rm s}$ for $m_{\rm ref} = 60{\rm M}_\odot$ \citep{2006A&A...457.1015H, 2007ARA&A..45..177C,2017RSPTA_37560269V, 2020MNRAS.499..873S}, 
	$L_{\rm ref} = 2.5\cdot 10^{38}{\rm erg}/{\rm s}$ for $m_{\rm ref} = 120{\rm M}_\odot$ \citep{2011A&A...530L..14B}, 
	at solar metallicity.
	We calibrated those and the power exponents to ensure continuity at the breakpoints, while satisfying the general slope trend -- OB stars follow a steeper profile than WR stars \citep{2001A&A...369..574V,2004A&A...415..349R}; for the very massive stars ($> 120 {\rm M}_\odot$) the curve flattens \cite{2020MNRAS.499..873S} -- and respecting theoretical and observational limits \citep{2001A&A...369..574V,2017A&A...603A.118R}. 
	Finally, we adopted a single, population-averaged, metallicity exponent factor, $f_Z = 0.8$, for all stars of $m \geq 8{\rm M}_\odot$ \citep{2001A&A...369..574V,2007ARA&A..45..177C,2017A&A...603A.118R}. This average exponent is sufficient for our main purpose, which is to get an approximate estimate of the gas depletion timescale, capturing the essential property that massive stars' wind energy depends strongly on metallicity.
	
	The wind energy is subject to losses and redistribution inside the cluster \citep{2013A&A...550A..49K,2014MNRAS.442..694D}.
	We adopted a wind feedback efficiency of $0.1$ \citep{2003ApJ...590..791S,2013A&A...550A..49K,2014MNRAS.442..694D,2015MNRAS.448.3248G}. 
	This includes not only energy loss (radiative cooling, mixing, leakage, incomplete coupling, and so on), but also the fraction of energy that is distributed inefficiently  (translational energy converted to rotational --turbulence -- geometry, and so on), so that only the energy that is actually used for gas expulsion is assumed to be counted.
	
	\subsection{SNe}\label{sec:SN}
	
	For core-collapse SNs (CCSNe), we set a lower $m_{\rm ZAMS}$ at $8 {\rm M}_\odot$ and a maximum possible $m_{\rm ZAMS}$ that increases with decreasing metallicity, particularly at $30{\rm M}_\odot$, $35{\rm M}_\odot$, and $40 {\rm M}_\odot$ for solar, subsolar ($\sim 0.1 Z_\odot$), and low ($\sim 0.01 Z_\odot$) metallicities, respectively (e.g., \citet{2002ApJ...567..532H, 2003ApJ...591..288H, 2006NuPhA.777..424N, 2016ApJ...821...38S}).
	The tabulated data of \cite{1995ApJS..101..181W} can be approximated, sufficiently for our purposes, with a simple second-order polynomial fit for solar, subsolar, and low metallicities, respectively, as
	\begin{equation}
		\begin{array}{ll}
			E_{\rm CCSN, sol} = \left(\frac{m_{\rm ZAMS} - 4.5}{7.3}\right)^2\cdot 10^{50}{\rm erg}, &
			8 \leq m_{\rm ZAMS}  \leq 30 ,
			\\
			E_{\rm CCSN, sub} = \left(\frac{m_{\rm ZAMS} - 3.0}{8.3}\right)^2\cdot 10^{50}{\rm erg}, &
			8 \leq m_{\rm ZAMS}  \leq 35 ,
			\\
			E_{\rm CCSN, low} = \left(\frac{m_{\rm ZAMS} - 2.6}{8.7}\right)^2\cdot 10^{50}{\rm erg}, &
			8 \leq m_{\rm ZAMS}  \leq 40 .
		\end{array}
	\end{equation}
	which give energies of $\sim (4-4.5)\cdot 10^{50}{\rm erg}$ for a $20{\rm M}_\odot$ star depending on the metallicity. 
	
	Regarding pulsational pair-instability SNe (PPISNe) and  pair-instability SNe (PISNe), we find that tabulated and plotted data of PPISNe \citep{2017ApJ...836..244W} and PISNe \citep{2002ApJ...567..532H, 2011ApJ...734..102K} are effectively modeled by simple linear fits.
	For PPISN we adjusted $E = 10^{50}{\rm erg}$ for $m_{\rm ZAMS} = 80$ and $E = 10^{51}{\rm erg}$ for $m_{\rm ZAMS} = 125$ \citep{2017ApJ...836..244W},
	\begin{equation}
		E_{\rm PPISN} = \frac{1}{5}(m_{\rm ZAMS} - 75)\cdot 10^{50}{\rm erg}, \quad
		75 \leq m_{\rm ZAMS}  \leq 135 .
	\end{equation}
	
	For PISN we adjusted $E = 5\cdot 10^{51}{\rm erg}$ for $m_{\rm ZAMS} = 140{\rm M}_\odot$ and $E = 5\cdot 10^{52}{\rm erg}$ for $m_{\rm ZAMS} = 230{\rm M}_\odot$ \citep{2002ApJ...567..532H, 2011ApJ...734..102K},
	\begin{equation}
		E_{\rm PISN} = \frac{1}{2}(m_{\rm ZAMS} - 130)\cdot 10^{51}{\rm erg}, \quad
		135 \leq m_{\rm ZAMS}  \leq 230 .
	\end{equation}
	We assumed that PISNe can occur only at low metallicity, while PPISN occur at both subsolar and low metallicity. 
	
	Only a fraction of the total energy generated by a SN explosion contributes efficiently to the depletion of the ambient gas \citep{2013A&A...550A..49K,2019MNRAS.488.3376P,2023A&A...676A..67W}.
	We adopted a SN feedback efficiency equal to $0.1$ in agreement with recent observations \citep{2023A&A...676A..67W}.
	
	We assumed a Kroupa piecewise initial mass function (IMF) for the stars \citep{kroupa2002} with slopes of $0.3$, $1.3$, and $2.3$ for the low-mass ($0.01 \leq m < 0.08$), intermediate-mass ($0.08 \leq m < 0.5$), and high-mass ($0.5 \leq m < m_{\rm max}$) regimes, respectively. We assumed $m_{\rm max} = 135 {\rm M}_\odot$ for the solar and subsolar metallicity cases and $m_{\rm max} = 230 {\rm M}_\odot$ for the low-metallicity case.
	We calculated the stellar lifetimes, extrapolating the tabulated data of \cite{2012A&A...537A.146E} for solar metallicity, the data of \cite{2013A&A...558A.103G} for subsolar metallicity ($Z = 0.002$), and the data of \cite{2019A&A...627A..24G} for low metallicity ($Z = 0.0004$).

	\begin{figure}[!htbp]
		\centering
		\begin{subfigure}[b]{0.85\columnwidth}
			\centering
			\includegraphics[width=0.9\textwidth]{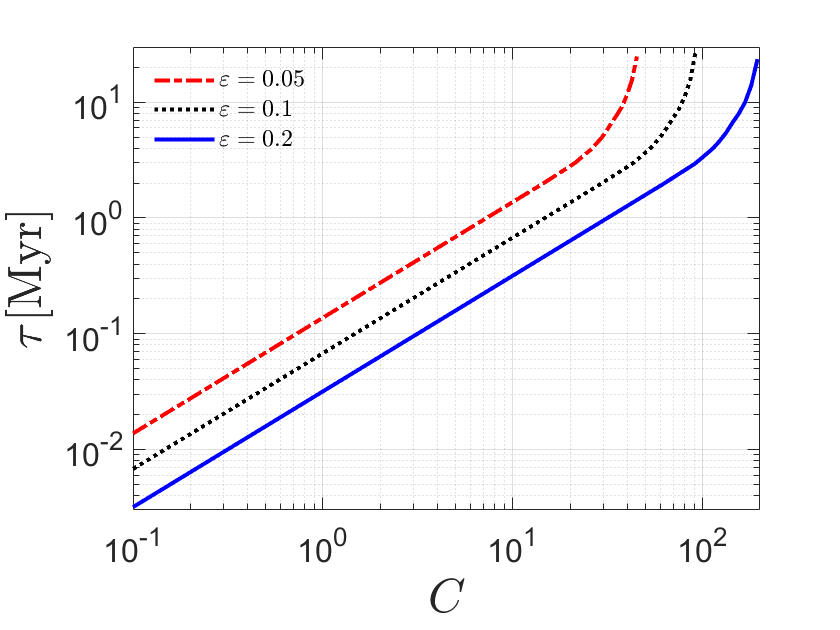}
			\caption{Solar metallicity.}
			\label{fig:tau_sol}
		\end{subfigure}
		\begin{subfigure}[b]{0.85\columnwidth}
			\centering
			\includegraphics[width=0.9\textwidth]{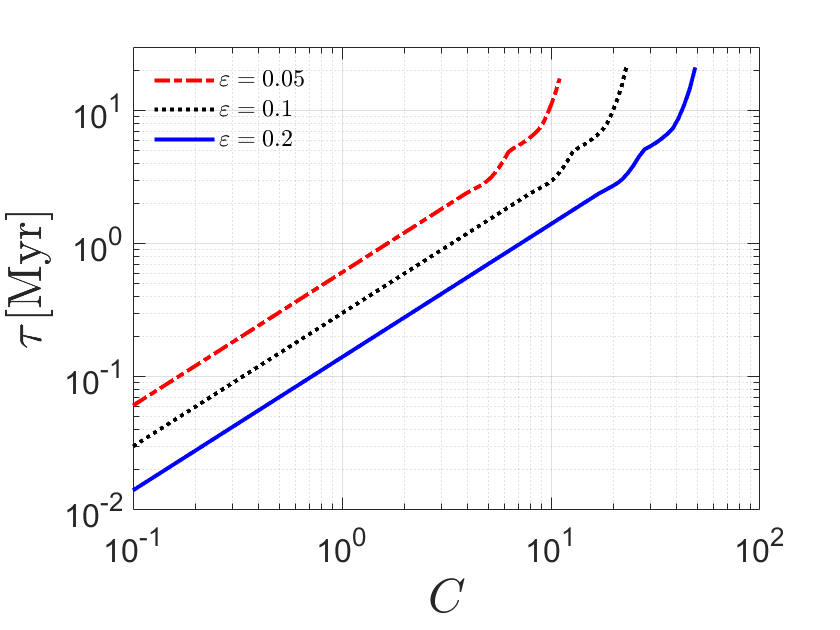}
			\caption{Subsolar metallicity.}
			\label{fig:tau_sub}
		\end{subfigure}
		\begin{subfigure}[b]{0.85\columnwidth}
			\centering
			\includegraphics[width=0.9\textwidth]{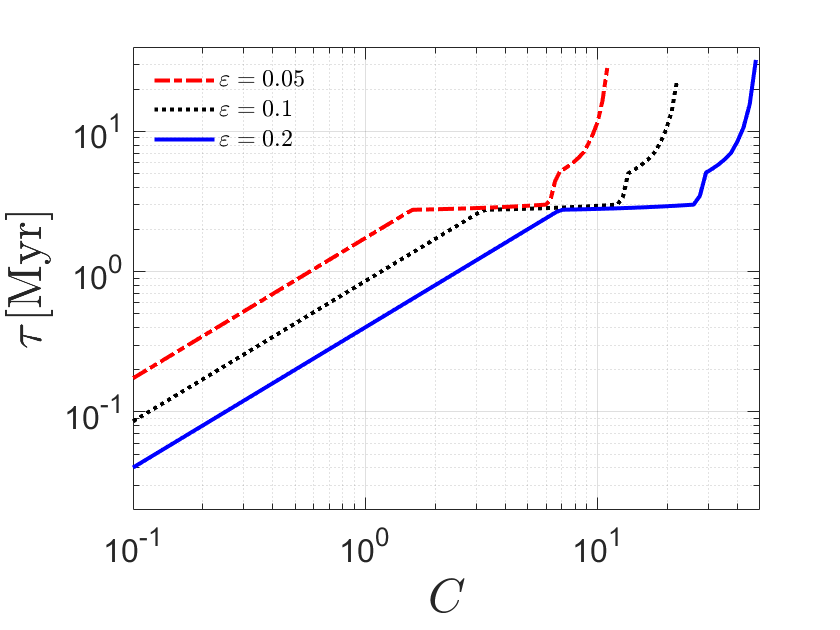}
			\caption{Low metallicity.}
			\label{fig:tau_low_PISN}
		\end{subfigure}
		\caption{Our estimated depletion timescale, $\tau$, with respect to the compactness, $C$, of a gaseous stellar cluster of any total mass. We considered three possible star formation efficiencies, $\varepsilon$. Each sub-figure $(a)$, $(b)$, $(c)$ corresponds, respectively, to solar, subsolar ($\sim 0.1 Z_\odot$), and low ($\sim 0.01 Z_\odot$) metallicity. 
			We assume that at subsolar metallicity PPISN operates, and that in the low-metallicity case both PPISN and PISN do.}
		\label{fig:tau_dep}
	\end{figure}
	
	\subsection{Depletion timescale estimation}
	
	The binding energy, $U \propto M^2 / R \sim C \cdot M$, grows linearly with the total mass of the cluster, $M$, for a fixed initial compactness, $C$, defined as
	\begin{equation}
		C \equiv \frac{M/10^5{\rm M}_\odot}{r_{\rm h}} .
	\end{equation}
	The feedback energy, on the other hand, grows as $E_{\rm feed} \propto M$ regardless of the compactness. Therefore, the effectiveness of gas depletion (depletion timescale) depends solely on the compactness, $C$, and of course the star formation efficiency, $\varepsilon$, not on the mass of the cluster. This basic energy scaling also suggests that for all clusters of any mass there should exist some upper compactness, above which gas expulsion is ineffective (see also \cite{2016A&A...587A..53K}). 
	
	The star formation efficiency is the other sensitive factor affecting the depletion timescale. Observational evidence at high redshift suggests a range of
	$\varepsilon \approx 0.05 - 0.3$ \citep{Inayoshi_2022}. At $z\approx 1$ observations suggest a typical value in giant molecular clouds of $\varepsilon \approx 0.1$ \citep{2019NatAs...3.1115D, 2023MNRAS.519.6222D}. We consider in our analysis the range $\varepsilon = 0.05 - 0.35$.

	In Figure \ref{fig:tau_dep} we depict the depletion timescale with respect to compactness for several $\varepsilon$ values and metallicities. We assume that in subsolar metallicity, Figure \ref{fig:tau_sub},  there occur PPISNe, but not an energetic PISN. At low metallicity, Figure \ref{fig:tau_low_PISN}, we consider that PISN do take place.
	
	Notice that for solar metallicity, although we consider the case that no PPISN and PISN operate, the timescale is shorter with respect to similar $\varepsilon$ and $C$ than in the subsolar and low-metallicity cases, because stellar winds are more severe. For the subsolar and low-metallicity cases, compactness values of $C\gtrsim 10$ have depletion timescales of $\tau \gtrsim 3{\rm Myr}$, high enough for BHs from sufficiently massive stars be generated early enough to accrete gas before it gets depleted, as we verify in detail below. Particularly significant is the plateau in the low-metallicity case at $\tau \simeq 3{\rm Myr}$ for $2 \lesssim C\lesssim 30$ . It signifies the ignition of PISN. These energetic SNs do not allow for the depletion timescale to increase even if compactness of the cluster is increased and/or stellar formation is decreased. This overlaping of $\tau$ for several $\varepsilon$ and $C$ values allows for larger $\varepsilon$ values to be as, or more, efficient regarding black hole mass function (BHMF) shift as lower $\varepsilon$ values, because higher $\varepsilon$ correspond to less diluted gas density profiles. We calculate this in detail below.
	
	\section{Physical model}\label{sec:model}
	
	For a given compactness, total cluster mass, star formation efficiency, and metallicity, we calculated the gas depletion timescale, $\tau$, generated a BH population with each BH generated at a different time depending on the ZAMS progenitor mass, and evolved these BHs in background gas and density Plummer profiles, taking into account the expansion of the cluster due to the mass loss. 
	We integrated the BHs equations of motion and accretion rate using a standard drift-kick scheme, with the deterministic forces -- gravity and dynamical friction -- evolved using an adaptive Runge-Kutta method and stochastic velocity increments, due to stellar perturbations and gas turbulence, applied at adaptive timesteps. We applied an additional recoil kick to the BH remnants of CCSN. This kick drives the majority of these types of BHs out of the cluster.
	
	We assumed that BH close encounters are implicitly considered in the stochastic stellar kicks calculation, i.e., we did not consider the BHs' mutual gravity separately. This allowed us to investigate a vast range of physical parameters, saving numerical time. The self-gravity of the BH population would be important if one calculated the formation of binary BHs and for significantly larger timescales, which we do not do here. In any case, the BH population contributes negligible gravity with respect to the stellar population and in addition each BH enters the simulation in its own moment of time. 
	In numbers, for example, a $10^6{\rm M}_\odot$ cluster generates $\sim 1000$ BHs, out of which $\sim 70\%$ escape at birth.
	The remaining $300$ BHs not only have negligible self-gravity with respect to the combined gas and stellar background but also each one enters the simulation in its own moment of time (depending on progenitor's ZAMS mass) over a range of $\sim 10 {\rm Myr}$ when our simulations approximately terminate. Also, out of the $300$ BHs, a fraction, $\sim 50$ BHs, do undergo a significant mass increase.
	
	Since the stellar feedback processes that expel the gas generate energy proportional to the cluster mass, it is natural to consider an exponential law, $\dot{M}_{\rm gas} = -M_{\rm gass}/\tau$, of gas depletion \citep{2007MNRAS.380.1589B}.
	We additionally assumed that the gas loss due to accretion onto BHs is already implicitly considered in this exponential law of gas depletion. This is justified as long as the total accreted mass is negligible with respect to the gas reservoir and certainly less than the remaining gas when we terminate the simulation. In practice, for all the simulations we present here the BHs do not accrete more than $\lesssim 0.5\%$ of the initial gas reservoir, while we terminated the simulation when exactly $99\%$ of the initial gas had depleted. 
	
	We assumed, furthermore, that the cluster expands according to the well-known adiabatic invariant \citep{1980ApJ...235..986H}
	\begin{equation}
		\frac{R(t)}{R_0} = \frac{M_0}{M(t)}.
	\end{equation}
	This assumes that the system remains nearly in virial equilibrium during gas depletion. This is sufficient for our purposes and is a good approximation of the $\tau \gtrsim 3{\rm Myr}$ cases that we consider here, improving the efficiency of our numerical code, while taking into consideration cluster dilution due to mass loss.
	
	The gas profile should naturally be more extended than the stellar component (e.g., \cite{2015ApJ...806...35J,2019Natur.569..519K}). We assumed that 
	\begin{equation}
		r_{{\rm h, gas}} = \left(\frac{1-\varepsilon}{\varepsilon}\right)^{b} r_{{\rm h, stars}}.
	\end{equation}
	The $b=1$ would correspond to both components being equally compact. We calibrated $b$ such that $r_{{\rm h, gas}} = 5\, r_{{\rm h, star}}$ for $\varepsilon = 0.1$ \citep{2013A&A...549A.132P}, which gives $b = 0.73$ and corresponds to a gaseous component that is less dense than the stellar component ($\rho_{\rm h, gas} = 0.07 \rho_{\rm h, stars}$) and significantly more extended.
	
	The gas temperature evolves as the cluster loses mass. We modeled the temperature evolution with a typical logarithmic scaling \citep{2003adu..book.....D},
	\begin{equation}
		T_0(t) = T_{\rm ini} - (T_{\rm ini} - T_{\rm fin}) \frac{\ln(M(t)/M_0)}{\ln(M_{\star}/M_0)}
		,\end{equation}
	where $T_0(t)$ is the central temperature with respect to time, and we set $T_{\rm ini} = 1.2 \times 10^4$ K the initial central temperature, $T_{\rm fin} = 4 \times 10^3$ K the final central temperature when gas is $99\%$ depleted. We further denote $M(t) = M_{\star} + M_{\rm gas}(t)$ the total cluster mass, and $M_0$ the initial total cluster mass.
	The thermal central velocity dispersion is related to temperature through
	$\sigma_{0,T}(t)^2 = k_B T_0(t)/m_{\rm eff}$,
	where $m_{\rm eff} = 0.6 \, m_p$ is the effective particle mass for the ionized gas, and we assume $\sigma_T(r,t)$ follows the Plummer profile with $\sigma_T(r=0,t) = \sigma_{0,T}(t)$.
	The local sound speed is then directly given by
	$c_s^2(r,t) = \gamma \sigma_{T}^2(r,t)$,
	where $\gamma = 5/3$ is the adiabatic index.
	
	\begin{figure}[!htbp]
		\centering
		\includegraphics[width=0.8\columnwidth]{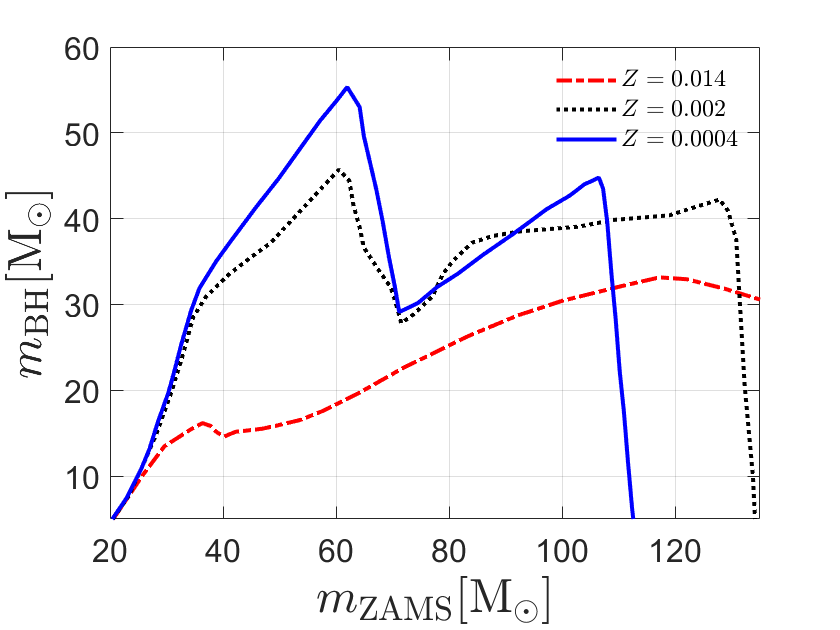}
		\caption{BH-remnant mass with respect to the ZAMS progenitor's mass, reproduced from \cite{2017MNRAS.470.4739S}, used here in determining the BH entry time in our simulation.}
		\label{fig:mass_BH_ZAMS}
	\end{figure}
	
	\subsection{BH generation}
	
	We generated the BH population stochastically, assigning as the BH IMF a Salpeter mass function \citep{2019ApJ...878L...1P} with a power exponent equal to $2.37$. We want to evaluate the shift of the BH IMF to higher mass values, which is why we want to have a definite power exponent for the BH IMF, which we can directly compare to the shifted BHMF after the gas gets depleted (at our fixed $99\%$ threshold).
	For the BH IMF, we assumed a minimum BH mass of $5{\rm M}\odot$ and a maximum BH mass that depends on the metallicity; that is, $55 {\rm M}\odot$ for low metallicity, $45 {\rm M}\odot$ for subsolar metallicity, and $33 {\rm M}\odot$ for solar metallicity. These values were extracted from the zero-age main sequence (ZAMS) progenitor versus BH mass figure 2 of \cite{2017MNRAS.470.4739S}, which includes PPISN and PISN. For completeness, we reproduce this data here, in our Figure \ref{fig:mass_BH_ZAMS}.
	We assigned a ZAMS mass to each BH based on Figure \ref{fig:mass_BH_ZAMS} and if several possible ZAMSs existed we also assigned a probability assuming a stellar Kroupa IMF as is described in section \ref{sec:SN}.
	Then, each BH entered the simulation at its own moment of time equal to the lifetime of the ZAMS progenitor assigned to it. 
	
	We further categorized the BHs depending on their formation type. If the ZAMS mass lay in the CCSN range we assigned the CCSN-BH type, if it lay in the range between the maximum CCSN and minimum PPISN thresholds we assigned it the direct-collapse-BH type, and if it lay in the PPISN range we assigned the type PPISN-BH. However, there are some additional considerations stemming from the Figure \ref{fig:mass_BH_ZAMS}. We can see that for subsolar and low metallicity there are two peaks. The first peak signifies the minimum possible mass loss during direct collapse. Between the peaks stars enter the PPISN regime where multiple pulsations eject significant envelope mass. At the second peak, stars eject the minimum possible mass through pulsations to drop below the pair-instability threshold, still allowing the remaining core to collapse into a BH. Beyond some ZAMS mass value complete PISN occurs, totally disrupting the star with no remnant. Therefore, we further subcategorized the direct-collapse-BHs born from ZAMS on the left part of the first peaks as direct-collapse-BHs from low ZAMS and those from the valley between the peaks as direct-collapse-BHs from high ZAMS. Likewise, we have PPISN-BHs from low ZAMS for those generated from the valley between the peaks and PPISN-BHs from high ZAMS for those generated from the right part of the second peak.
	
	We assumed that the massive stellar progenitors had undergone mass segregation and reached Brownian equilibrium within the stellar profile \citep{2021A&A...646A..20R}. Consequently, we sampled the initial BH position from the \cite{2021A&A...646A..20R} profile, $\rho_{\rm BH}(r) \propto \sigma_{\star}(r)^{-2} \cdot \exp \left\lbrace - \int \frac{m_{\rm BH}}{m_{\star} \sigma_{\star}(r)^2} d\Phi_{\star}(r) \right\rbrace$. We sampled the iniital BH velocities from a Maxwell distribution using Jeans theorem for this density profile.
	
	Black holes formed by direct collapse, in the absence of energetic mass loss, are not expected to receive significant kick velocities \citep{2022NatAs...6.1085S}. This is also the case for PPISN-formed BHs. The PPISN pulses expel most of the envelope prior to core collapse, so that the final core implosion proceeds as an almost spherically symmetric direct-collapse event, with the resulting BH receiving at most a few kilometers per second kicks \citep{2022MNRAS.512.4503R}. We therefore assumed that direct-collapse-BHs and PPISN-BHs attain the kicks' magnitude, at most comparable to the 1D rms velocity of the stellar progenitor, and therefore that the kicks' contribution to the initial BH velocity is implicitly considered in our random intial BH-velocity generator. On the other hand, the CCSN-BHs are expected to receive significant natal kicks due to asymmetries in the explosion (e.g., \cite{2016MNRAS.456..578M}). The CCSN remnants are reasonably expected to attain kick velocities that follow a Maxwellian \citep{2005MNRAS.360..974H}. For CCSN-BHs we adopted a prescription similar to \cite{2020ApJ...891..141G} for the 1D rms velocity of the natal kicks, $\sigma_{\rm NK}$, of the Maxwellian distribution; that is, $\sigma_{\rm NK} \propto m_{\rm BH}^{-1}$. We calibrated it to account for the possibility of kicks as high as $100{\rm km}/{\rm s}$ \citep{2017PhRvL.119a1101O,2019MNRAS.489.3116A}, assuming $\sigma_{\rm NK} = 50 {\rm km}/{\rm s} \cdot (m_{\rm BH}/10)^{-1}$ for CCSN-BHs.

	\subsection{Gas accretion}
	
	In the dense, hot, turbulent gaseous environment that we considered, and given the BH motion at speeds comparable to but not significantly larger than the speed of sound, $c_s$, it is justified to consider an isotropic hot-type accretion, whose typical representative is Bondi-Hoyle accretion \citep{Merritt_book}.
	The spherically symmetric accretion rate, for a cross section of $\pi R_\text{acc}^2$, 
	is \citep{1985Sci...230.1269F}
	\begin{equation}\label{eq:dm}
		\dot{m}_{\rm feed} = \pi \rho_\text{gas} v_\text{rel} R_\text{acc}^2
		,\end{equation}
	where the BH relative velocity is
	\begin{equation}
		v_\text{rel} = \sqrt{v^2 + c_s^2}.
	\end{equation} 
	
	The proper accretion radius inside a gaseous stellar cluster depends on the relative amount of gas and on the radial position of the accretor \citep{2001MNRAS.324..573B}.
	When gas is less dominant, the appropriate accretion radius is the Bondi-Hoyle radius, $R_{\rm B}$:
	\begin{equation}
		R_\text{B} = \frac{2 G m_{\rm BH}}{v_\text{rel}^2}.
	\end{equation}
	However, the accretion rate is determined by a tidal-lobe accretion radius, when the gas dominates the cluster potential \citep{1971ARA&A...9..183P},
	\begin{equation}
		R_\text{tid}(r_i) \sim 0.5 \left( \frac{m_{\rm BH}}{M(r<r_i)} \right)^{1/3} r_i,
	\end{equation}
	where $M(r<r_i)$ is the total mass of the cluster within the radial position, $r_i$, of the $i$th BH. 
	We chose the accretion radius to be the smaller of the two at any instant in time \citep{2001MNRAS.324..573B},
	\begin{equation}\label{eq:R_acc}
		R_{\rm acc}(t) = \min \left\lbrace R_{\rm B}, \, R_{\rm tid} \right\rbrace
		,\end{equation}
	adopting the most conservative assumption. 
	
	\cite{2016MNRAS.459.3738I} have shown that for $1 < \dot{m}_{\rm feed} / \dot{m}_{\rm Edd,0} < 100$ the mean accretion rate cannot exceed $\left\langle \dot{m}_{\rm acc} \right\rangle \lesssim 10 \dot{m}_{\rm Edd,0}$.
	The quantity
	$
	\dot{m}_{{\rm Edd},0} = 4\pi G m_{\rm BH} m_p/\eta_0 \sigma_T c 
	$
	is the reference Eddington rate value for the radiative efficiency $\eta_0 = 0.057$. We therefore adopted a maximum threshold,
	\begin{equation}
		\dot{m}_{\rm acc} = \min\left\lbrace \dot{m}_{\rm feed}, 10 \dot{m}_{{\rm Edd},0} \right\rbrace. 
	\end{equation}
	This is a conservative assumption, especially at low metallicity and for thick disks. \cite{2016MNRAS.459.3738I} report that though the mean accretion rate cannot exceed a few times the $\dot{m}_{{\rm Edd},0}$ (when $\dot{m}_{\rm feed} < 100\dot{m}_{{\rm Edd},0}$), there occur episodic bursts of accretion rate comparable to $\dot{m}_{\rm feed}$. In this regime the accretion rate is not strictly steady. We adopted a constant cap as a simple, conservative assumption. In practice, in the physical conditions we consider here, the $\dot{m}_{\rm feed}$ we find is less than $\lesssim 10{\rm M}_\odot / {\rm Myr}$ for a $50{\rm M}_\odot$ BH, close to the Eddington accretion rate for a thick disk radiative efficiency, $\eta \sim 0.01$, which is about $5\dot{m}_{{\rm Edd},0}$, i.e., below our applied threshold. The $\dot{m}_{\rm feed}$ tends to attain values above our threshold when BHs approach masses of $\sim 10^3{\rm M}_\odot$ in the physical conditions we consider here.
	
	A concern might be that CCSN and PPISN events form dilute bubbles that expand rapidly around the progenitor star, and therefore whether the accretion rates we consider are appropriate for CCSN-BHs and PPISN-BHs. In our physical conditions, however, a BH formed from a CCSN or PPISN has sufficient time to escape the bubble, while the bubble itself has sufficient time to refill with gas. Specifically, the bubble's maximum radius is given by the stalling formula $R_{\rm stall} \simeq 0.6(E_{\rm SN}/\rho c_s^2)^{1/3}$ 
	\citep{1987ApJ...317..190M}, which is reached at time $t_{\rm stall} = 0.4 R_{\rm stall} / c_s$ \citep{1946JApMM..10..241S,1950RSPSA.201..192T}, assuming that the shock stalls when $v_{\rm shock} \approx c_s$. The BH requires time $t_{\rm cross} \sim R_{\rm stall}/v_{\rm BH}$ to cross the bubble. The bubble refills on a timescale of $t_{\rm refill} \sim R_{\rm stall}/c_s$. For $E_{\rm SN}\sim 10^{51}-10^{52} \, {\rm erg}$, all of these timescales are on the order of $1-10 \, {\rm kyr}$. This is short compared to the timescale on which the BHMF-shift operates ($\sim 10 \, {\rm Myr}$). Even heavy BHs that are of low mobility have the opportunity to accrete gas because at these high temperatures the bubble refills rapidly relative to the accretion process timescale. Nevertheless, local conditions may stochastically suppress accretion for CCSN-BHs and PPISN-BHs. Therefore, our results are expected to be less accurate for CCSN-BHs and PPISN-BHs than for direct-collapse-BHs, which form without disrupting the gas in their immediate vicinity.
	
	\subsection{Dynamical friction}
	
	The deterministic forces applied to the BHs are the gravitational force from the stellar and gaseous background and the dynamical friction due to both components. We used the standard Chandrasekhar dynamical friction formula to model the deceleration of BHs due to gravitational interactions with surrounding stars:
	\begin{equation}\label{eq:a_df_star}
		\vec{a}_{\text{DF},{\star}} = -4\pi G^2 m_{\rm{BH}} \rho_{\star}(r) \ln\Lambda \frac{\vec{v}}{v^3} \left( \text{erf}(X) - \frac{2X}{\sqrt{\pi}} e^{-X^2} \right)
		,\end{equation}
	where $X = \frac{v}{\sqrt{2}\sigma_{\star}}$, with $\sigma_{\star}$ the local stellar velocity dispersion.
	We denoted as $v$ the velocity of the BH, and the Coulomb logarithm is $\ln\Lambda = \ln (b_\text{max}/b_\text{min})$, with
	$b_\text{max} = r_{\rm cs}$ and $b_\text{min} = G m_{\rm BH}/3v_{\rm rel}^2$,
	where $r_{\rm cs}$ is the stellar Plummer radius.
	
	In a gaseous medium, the dynamical friction formula needs modification to account for the hydrodynamic nature of the interaction \citep{1999ApJ...513..252O}
	\begin{equation}
		\vec{a}_{\text{DF,gas}} = -4\pi G^2 m_{\text{BH}} \rho_{\text{gas}}(r) \frac{\vec{v}}{v^3} \mathcal{F}(v, c_s)
		,\end{equation}
	where $c_s$ is the sound speed in the gas and $\mathcal{F}$ is a function that accounts for the Mach number, $\mathcal{M} \equiv v/c_s$, dependence
	\begin{equation}
		\mathcal{F}(v, c_s) = 
		\begin{cases}
			\frac{1}{2} \ln \left( \frac{1 + \mathcal{M}}{1 - \mathcal{M}}\right) - \mathcal{M}, & \text{if } \mathcal{M} < 1 \\
			\frac{1}{2} \ln \left( 1 - \frac{1}{\mathcal{M}^2}\right) + \ln\Lambda, & \text{if } \mathcal{M} > 1 ,
		\end{cases}
	\end{equation}
	with $\mathcal{F} \rightarrow \ln\Lambda$ in the singular limit $\mathcal{M} \rightarrow 1$.
	
	We added to this the deceleration due to the mass increase \citep{2009ApJ...696.1798T,2011MNRAS.416.3177L,2014A&A...561A..84L} 
	\begin{equation}\label{eq:a_acc}
		\bm{a}_\text{acc} = -\frac{\dot{m}_{\rm BH}}{m_{\rm BH}}\bm{v},
	\end{equation}
	which may be understood simply as manifesting angular momentum preservation \citep{2019A&A...621L...1R}.
	
	\begin{table}[!tbp]
		\caption{\label{tab:dm_fit} BH mass growth, $\Delta m$, with respect to the initial mass, $m_{0}$.}
		\centering
		\begin{tabular}{c | c c}
			& Subsolar $Z$ & Low $Z$
			\\
			\toprule
			\begin{minipage}{0.8in}
				$r_{\rm h} = 1.0 {\rm pc}$ \\
				$\varepsilon = 0.06$
			\end{minipage}
			&
			$\Delta m = \left(\frac{m_{0}}{22.4} \right)^{4.6}$
			&
			$\Delta m = \left(\frac{m_{0}}{25.0} \right)^{4.3}$
			\\
			\midrule		
			\begin{minipage}{0.8in}
				$r_{\rm h} = 0.7 {\rm pc}$ \\
				$\varepsilon = 0.1$
			\end{minipage}
			&
			$\Delta m = \left(\frac{m_{0}}{21.7} \right)^{6.0}$
			&
			$\Delta m = \left(\frac{m_{0}}{24.5} \right)^{5.4}$
			\\
			\midrule		
			\begin{minipage}{0.8in}
				$r_{\rm h} = 0.5 {\rm pc}$ \\
				$\varepsilon = 0.2$
			\end{minipage}
			&
			$\Delta m = \left(\frac{m_{0}}{25.0} \right)^{5.3}$
			&
			$\Delta m = \left(\frac{m_{0}}{27.0} \right)^{4.9}$
		\end{tabular}
		\tablefoot{Initial cluster's total mass, $M = 10^6{\rm M}_\odot$. The overall fit of $\Delta m$ for BHs of the main branch CCSN-BH and direct-collapse-BH, displayed in the panels of Figure \ref{fig:BHmass_scatter_M1.0e6} for the same, several $\varepsilon$ and $r_{\rm h}$ values.}
	\end{table}

	\subsection{Stellar stochastic kicks}
	
	The granularity of the stellar component induces local perturbations in the background mean gravitational field. We modeled these stochastic velocity perturbations on BHs through stellar encounters as a white noise process (e.g., \cite{2021A&A...646A..20R}) derived from the fluctuation-dissipation theorem
	\begin{equation}
		\delta \mathbf{v} = \sqrt{2 D_{\star} \, \Delta t} \, \boldsymbol{\xi}
		,\end{equation}
	\noindent where $\boldsymbol{\xi}$ is a 3D Gaussian random vector with zero mean and unit variance and $\Delta t$ is the timestep. We describe below our adaptive timestep formulation. The diffusion coefficient is
	\begin{equation}
		D_{\star} = \frac{\langle m_{\star} \rangle}{m_{\rm BH}} \, \eta_{\star} \, \sigma_{\star}^2 .
	\end{equation}
	The stellar dynamical friction coefficient, $\eta_{\star}$, follows from Chandrasekhar formula \eqref{eq:a_df_star}.
	The diffusion coefficient scales inversely with BH mass, ensuring that more massive BHs experience proportionally weaker stochastic velocity kicks. 
	
	\subsection{Turbulent gas stochastic kicks}
	
	Gas turbulence is modeled as an Ornstein-Uhlenbeck (OU) process acting on the BH through a time-correlated turbulent acceleration field. OU is a well-defined stochastic process with a finite autocorrelation timescale and can be used to excite turbulent motions in simulations \citep{1988CF.....16..257E,SCHMIDT2006353,2010A&A...512A..81F}. It is generated by the OU stochastic differential equation,
	\begin{equation}\label{eq:OU_SDE}
		d\boldsymbol{\alpha} = - \frac{\boldsymbol{\alpha}}{\tau_{\rm OU}} dt + \sigma_a \sqrt{\frac{2}{\tau_{\rm OU}}} \, d\mathbf{W}
		,\end{equation}
	where $d\mathbf{W}$ is a 3D Wiener process, representing independent Gaussian white noise with zero mean and variance $dt$ in each spatial component. 
	We set the acceleration variance as
	\begin{equation}
		\sigma_a = \frac{G M_{\rm gas}}{\sqrt{3} \, r_{\rm vir,gas}^2}
	\end{equation}
	and the correlation time depends on the turbulence Mach number, 
	$\tau_{\rm OU} = r_{\rm cg}/\mathcal{M} c_s$ if $\mathcal{M} > 1$ or $\tau_{\rm OU} = r_{\rm cg}/c_s$ otherwise.
	
	The OU equation originates from the Langevin description of a particle subject to both linear damping and stochastic forcing. This formulation naturally produces time-correlated fluctuations with the autocorrelation function $\langle \boldsymbol{\alpha}(t) \cdot \boldsymbol{\alpha}(t') \rangle = \sigma_a^2 e^{-|t-t'|/\tau_{\rm OU}}$, capturing the finite coherence time of turbulent eddies in the gaseous medium.
	
	The OU-SDE \eqref{eq:OU_SDE} can be integrated directly to get the turbulent acceleration 
	\begin{equation}\label{eq:a_OU}
		\boldsymbol{\alpha}_{\rm turb}(t + \Delta t) = \boldsymbol{\alpha}_{\rm turb}(t) e^{-\Delta t/\tau_{\rm OU}} + \sigma_a \sqrt{1 - e^{-2\Delta t/\tau_{\rm OU}}} \, \boldsymbol{\xi}
		,\end{equation}
	where $\boldsymbol{\xi}$ is a 3D Gaussian random vector with zero mean and unit variance.
	We can now integrate the OU acceleration \eqref{eq:a_OU} to get the corresponding velocity kick over timestep $\Delta t$, 
	\begin{align}
		\delta \mathbf{v} = &\boldsymbol{\alpha}_{\rm turb}(t) \tau_{\rm OU} (1 - e^{-\frac{\Delta t}{\tau_{\rm OU}}}) 
		\nonumber \\
		&+ \sigma_a \tau_{\rm OU} \left(4e^{-\frac{\Delta t}{\tau_{\rm OU}}} - e^{-2\frac{\Delta t}{\tau_{\rm OU}}} + 2\frac{\Delta t}{\tau_{\rm OU}} - 3\right)^{\frac{1}{2}} \, \boldsymbol{\xi}'
		,\end{align}
	where $\boldsymbol{\xi}'$ is an independent Gaussian random vector. 
	
	\begin{figure*}[!htbp]
		\centering
		\begin{tabular}{cc}
			Subsolar metallicity & Low metallicity
			\\
			\begin{subfigure}[b]{0.85\columnwidth}
				\centering
				\includegraphics[width=0.9\textwidth]{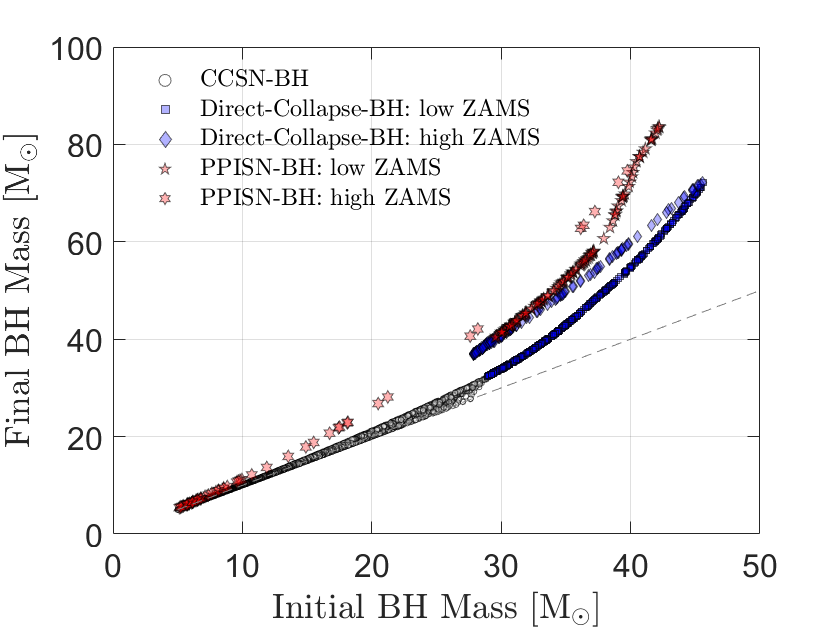}
				\subcaption{$r_{\rm h} = 1.0 {\rm pc}$, $\varepsilon = 0.06$
				}
				\label{fig:BHmass_scatter_M1.0e6_C10.0_eps0.06_Zsub}
			\end{subfigure}
			& 
			\begin{subfigure}[b]{0.85\columnwidth}
				\centering
				\includegraphics[width=0.9\textwidth]{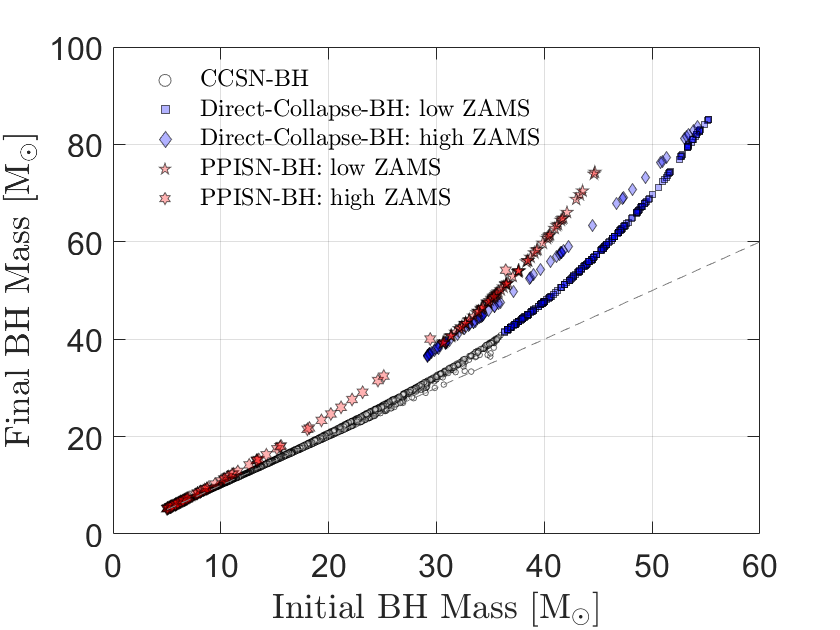}
				\subcaption{$r_{\rm h} = 1.0 {\rm pc}$, $\varepsilon = 0.06$
				}
				\label{fig:BHmass_scatter_M1.0e6_C10.0_eps0.06_Zlow}
			\end{subfigure}
			\\
			\begin{subfigure}[b]{0.85\columnwidth}
				\centering
				\includegraphics[width=0.9\textwidth]{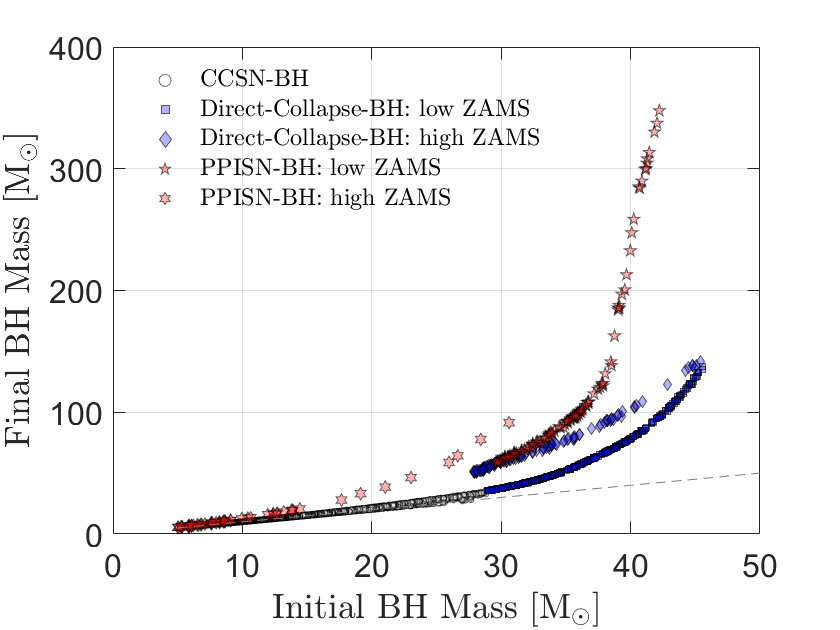}
				\subcaption{$r_{\rm h} = 0.7{\rm pc}$, $\varepsilon = 0.1$
				}
				\label{fig:BHmass_scatter_M1.0e6_C14.3_eps0.1_Zsub}
			\end{subfigure}
			& 
			\begin{subfigure}[b]{0.85\columnwidth}
				\centering
				\includegraphics[width=0.95\textwidth]{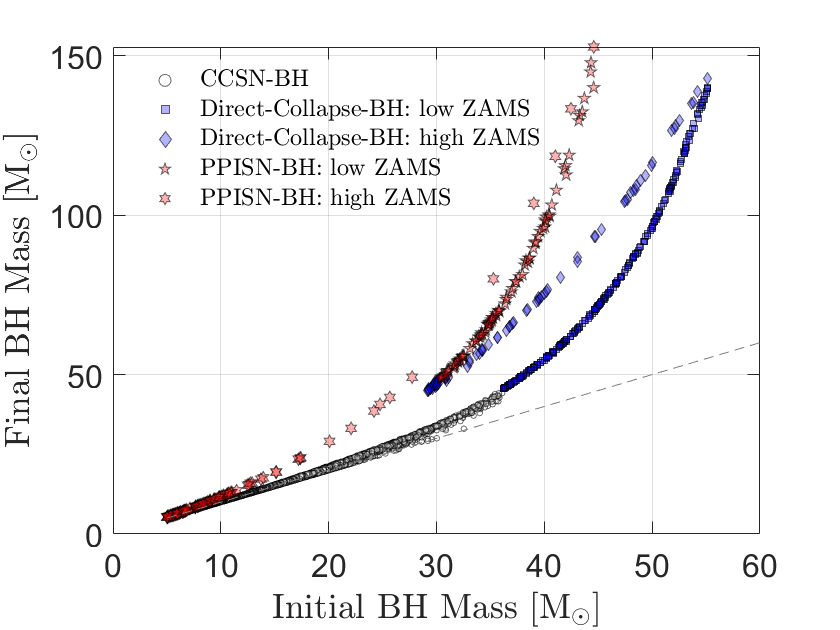}
				\subcaption{$r_{\rm h} = 0.7{\rm pc}$, $\varepsilon = 0.1$
				}
				\label{fig:BHmass_scatter_M1.0e6_C14.3_eps0.1_Zlow}
			\end{subfigure}
			\\
			\begin{subfigure}[b]{0.85\columnwidth}
				\centering			\includegraphics[width=0.95\textwidth]{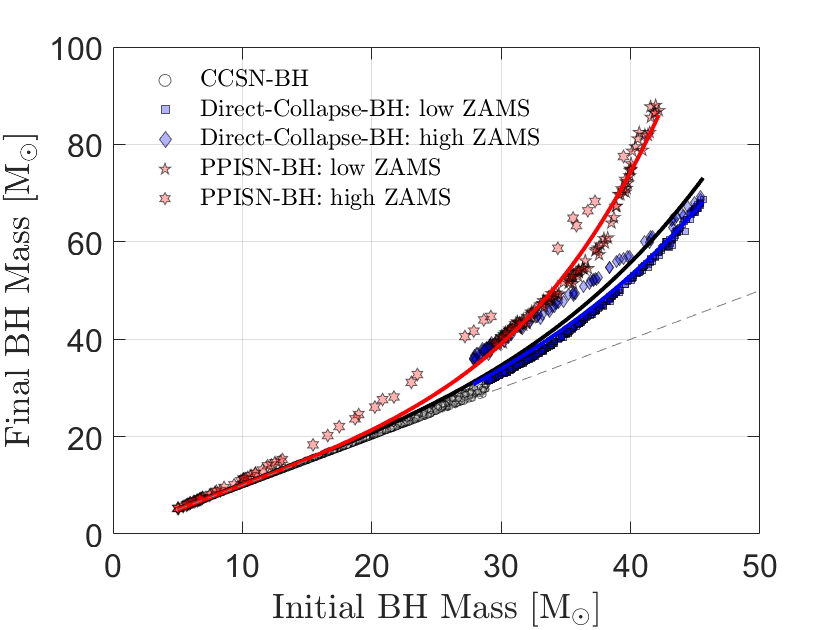}
				\subcaption{$r_{\rm h} = 0.5 {\rm pc}$, $\varepsilon = 0.2$
				}
				\label{fig:BHmass_scatter_M1.0e6_C20.0_eps0.2_Zsub}
			\end{subfigure}
			& 
			\begin{subfigure}[b]{0.85\columnwidth}
				\centering
				\includegraphics[width=0.95\textwidth]{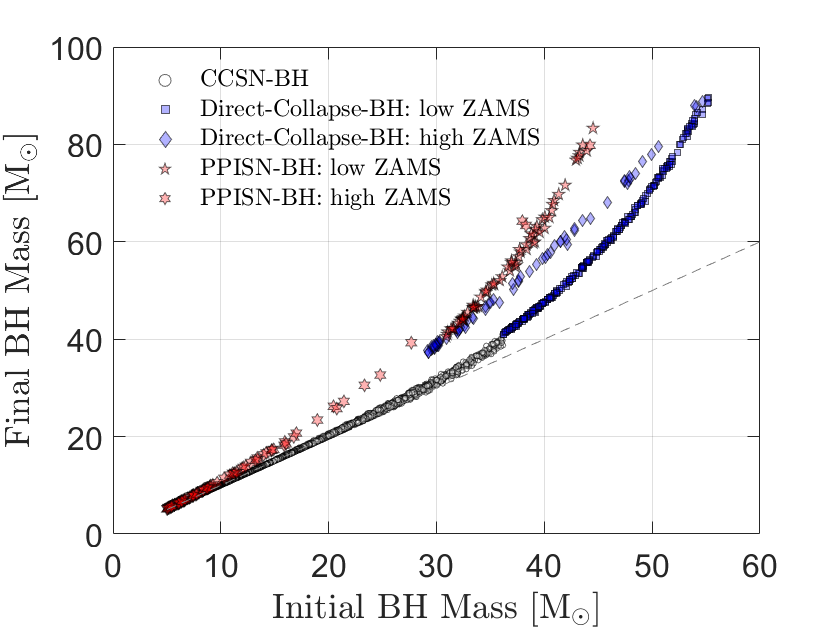}
				\subcaption{$r_{\rm h} = 0.5 {\rm pc}$, $\varepsilon = 0.2$
				}
				\label{fig:BHmass_scatter_M1.0e6_C20.0_eps0.2_Zlow}
			\end{subfigure}
		\end{tabular}
		\caption{Shifted BH masses, after stellar feedback expels  $99\%$ of the gas, with respect to the initial BH masses for a gaseous stellar cluster with total initial mass $M = 10^6{\rm M}_\odot$ containing $\sim 1000$ BHs, for several star formation efficiency and half-mass radius, i.e., compactness, values of the cluster. Each subfigure was generated by a set of ten simulations. The left column corresponds to subsolar metallicity ($\sim 0.1 Z_\odot$), the right column to low metallicity ($\sim 0.01 Z_\odot$). The solar metallicity case, not shown here, requires about double or higher compactness to have the same effect as the subsolar case.}
		\label{fig:BHmass_scatter_M1.0e6}
	\end{figure*}

	\begin{figure*}[!htbp]
		\centering
		\begin{tabular}{cc}
			Subsolar metallicity & Low metallicity
			\\	
			\begin{subfigure}[b]{0.85\columnwidth}
				\centering			\includegraphics[width=0.9\textwidth]{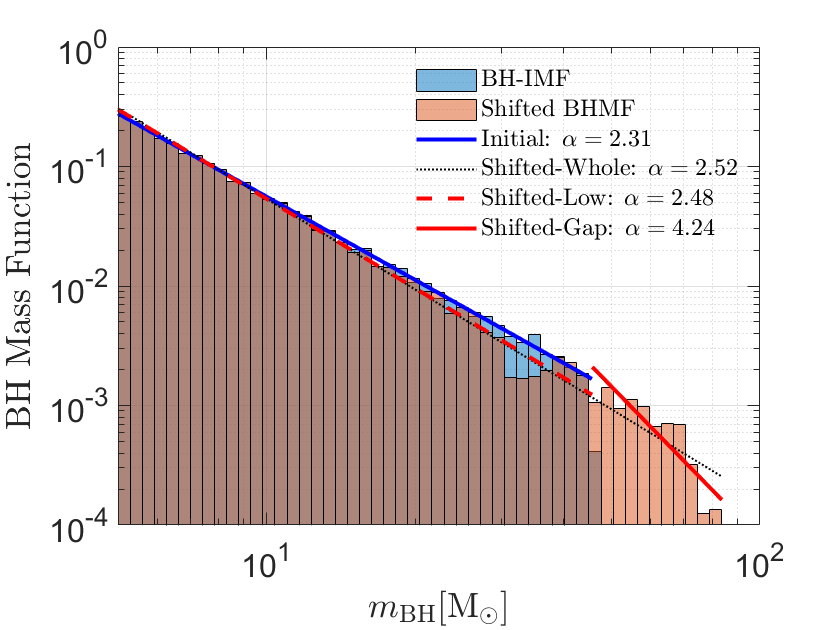}
				\subcaption{$r_{\rm h} = 1.0 {\rm pc}$, $\varepsilon = 0.06$
				}
				\label{fig:BHMF_M1.0e6_C10.0_eps0.06_Zsub}
			\end{subfigure}
			& 
			\begin{subfigure}[b]{0.8\columnwidth}
				\centering
				\includegraphics[width=0.95\textwidth]{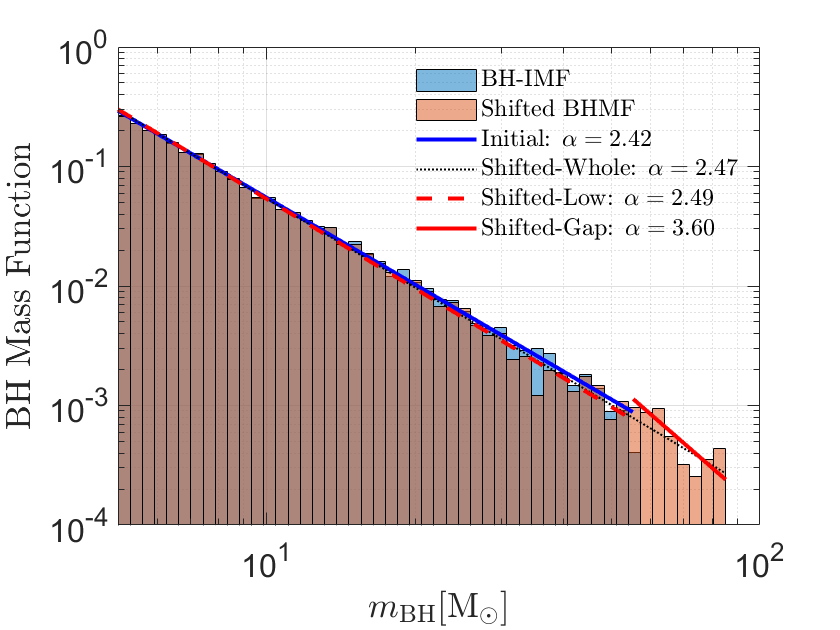}
				\subcaption{$r_{\rm h} = 1.0 {\rm pc}$, $\varepsilon = 0.06$
				}
				\label{fig:BHMF_M1.0e6_C10.0_eps0.06_Zlow}
			\end{subfigure}
			\\
			\begin{subfigure}[b]{0.85\columnwidth}
				\centering
				\includegraphics[width=0.9\textwidth]{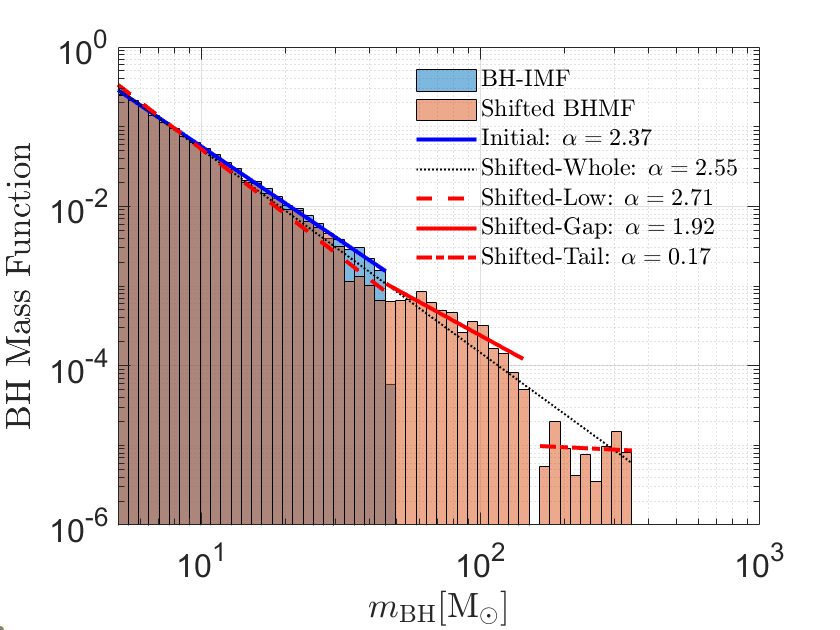}
				\subcaption{$r_{\rm h} = 0.7{\rm pc}$, $\varepsilon = 0.1$
				}
				\label{fig:BHMF_M1.0e6_C14.3_eps0.1_Zsub}
			\end{subfigure}
			&
			\begin{subfigure}[b]{0.85\columnwidth}
				\centering			\includegraphics[width=0.9\textwidth]{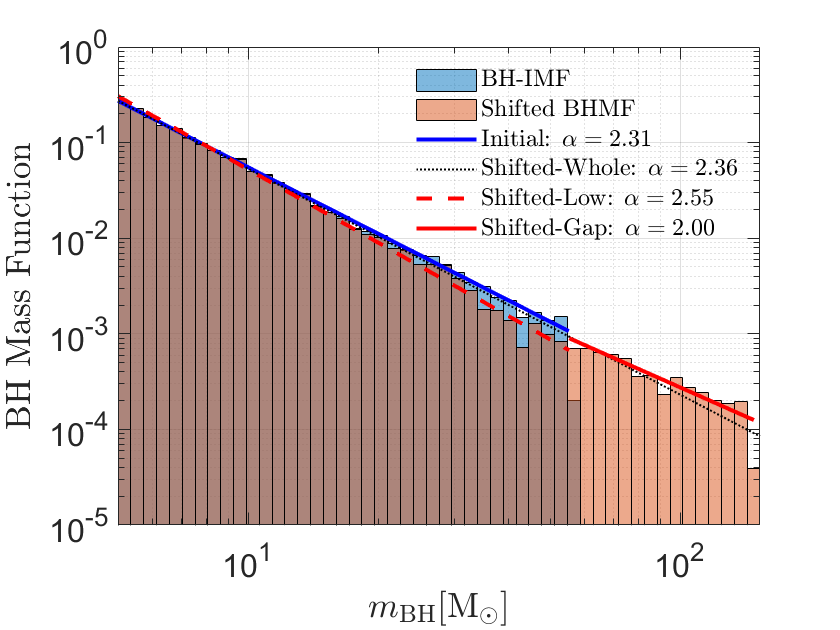}
				\subcaption{$r_{\rm h} = 0.7{\rm pc}$, $\varepsilon = 0.1$
				}
				\label{fig:BHMF_M1.0e6_C14.3_eps0.1_Zlow}
			\end{subfigure}
			\\
			\begin{subfigure}[b]{0.85\columnwidth}
				\centering
				\includegraphics[width=0.9\textwidth]{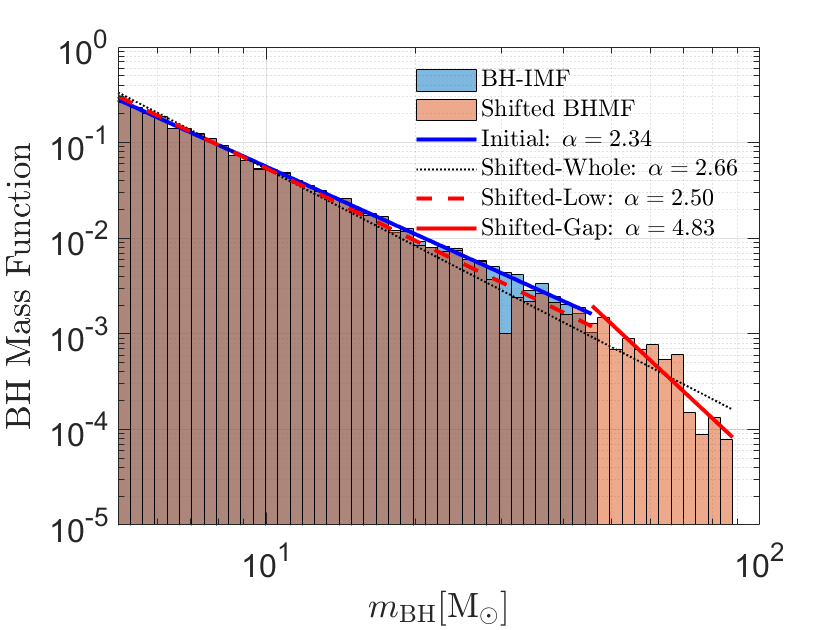}
				\subcaption{$r_{\rm h} = 0.5 {\rm pc}$, $\varepsilon = 0.2$
				}
				\label{fig:BHMF_M1.0e6_C20.0_eps0.2_Zsub}
			\end{subfigure}
			& 
			\begin{subfigure}[b]{0.85\columnwidth}
				\centering
				\includegraphics[width=0.9\textwidth]{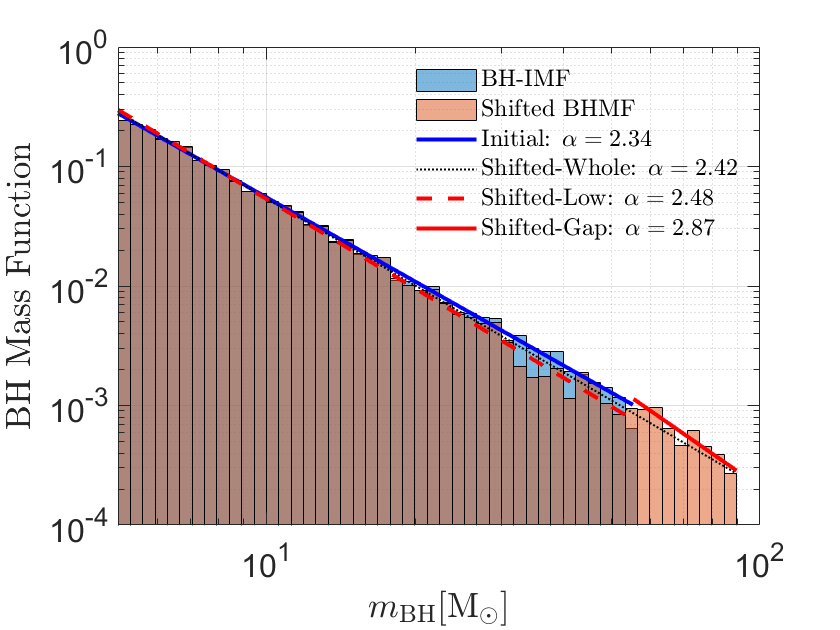}
				\subcaption{$r_{\rm h} = 0.5 {\rm pc}$, $\varepsilon = 0.2$
				}
				\label{fig:BHMF_M1.0e6_C20.0_eps0.2_Zlow}
			\end{subfigure}
		\end{tabular}
		\caption{BHMF shift, after stellar feedback expels  $99\%$ of the gas, for a gaseous stellar cluster as in Figure \ref{fig:BHmass_scatter_M1.0e6}. In the legend we depict the power exponent, $\alpha$, of several fits, namely of the BH IMF and of the shifted BHMF for several ranges; the whole range of values, the low-mass range up to the initial PISN cutoff, the range within the upper BH mass gap, and if applicable the tail of heavy BHs.}
		\label{fig:BHMF_M1.0e6}
	\end{figure*} 
	
	\subsection{Adaptive timestep of stochastic kicks}
	
	We adaptively determined the integration timestep of both stellar and gas stochastic kicks for each BH based on the minimum of relevant physical timescales locally to ensure numerical stability and accuracy,
	\begin{equation}
		\Delta t = f_{\rm dt} \cdot \min\left\{ \tau_{\rm grav}, \tau_{\rm cross}, \tau_{\rm df}, \tau_{\rm OU} \right\}
		,\end{equation}
	where $f_{\rm dt}$ is a safety factor with our default value of $f_{\rm dt} = 0.5$. We denote the dynamical timescale,
	$\tau_{\rm grav} = \sqrt{r^3/G M_{\rm encl}}$, the crossing timescale, $\tau_{\rm cross} = r_{\rm cs}/v$, the dynamical friction timescale, $\tau_{\rm df} = 1/(\eta_{\star} + \eta_{\rm gas})$, and the correlation time of the OU process, $\tau_{\rm OU}$.
	
	The timestep is further constrained by prescribed bounds,
	$\Delta t_{\rm min} \leq \Delta t \leq \Delta t_{\rm max}$,
	with default values of $\Delta t_{\rm min} = 10^{-5}{\rm Myr}$, $\Delta t_{\rm max} = 0.1{\rm Myr}$. In practice, these bounds are always satisfied.
	This approach is intended to ensure that the integration resolves the fastest relevant physical process while maintaining computational efficiency. We directly integrated the deterministic part -- gravity and dynamical friction -- between the stochastic timesteps using an adaptive Runge-Kutta method.
	
	\begin{figure*}[!htbp]
		\centering
		\begin{tabular}{cccc}
			& Solar metallicity & Subsolar metallicity & Low metallicity \\
			$10^5{\rm M}_\odot$ & \begin{minipage}[c]{0.28\textwidth}\centering\includegraphics[width=0.95\textwidth]{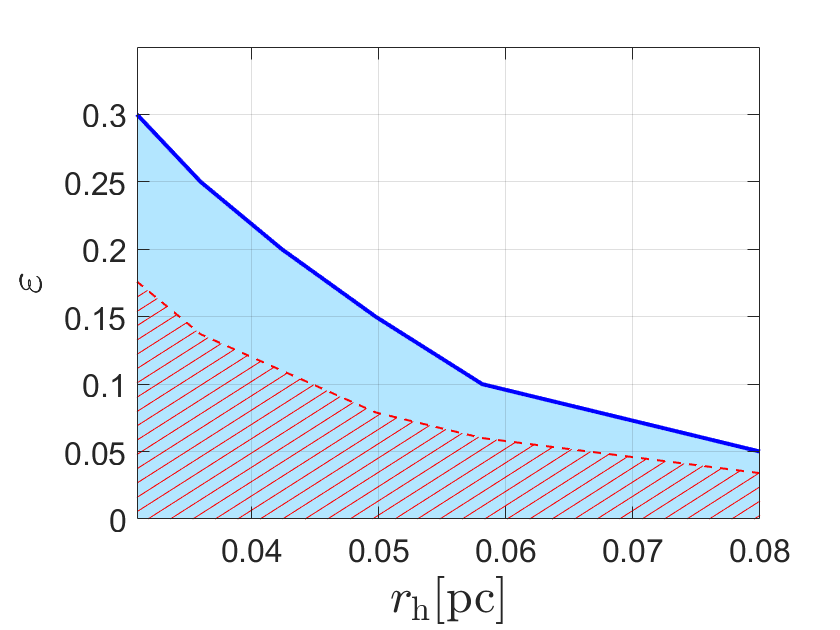}\end{minipage} & \begin{minipage}[c]{0.28\textwidth}\centering\includegraphics[width=0.95\textwidth]{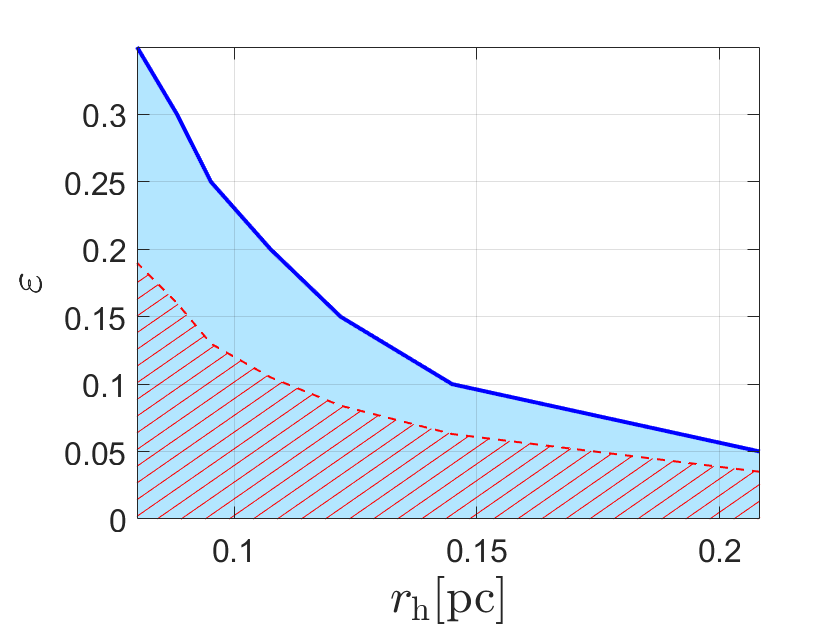}\end{minipage} & \begin{minipage}[c]{0.28\textwidth}\centering\includegraphics[width=0.95\textwidth]{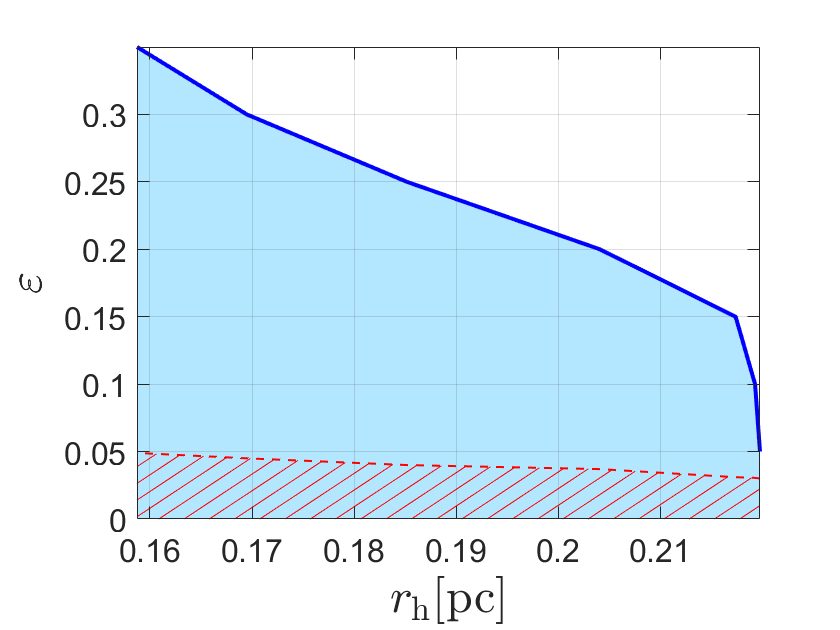}\end{minipage} \\
			$10^6{\rm M}_\odot$ & \begin{minipage}[c]{0.28\textwidth}\centering\includegraphics[width=0.95\textwidth]{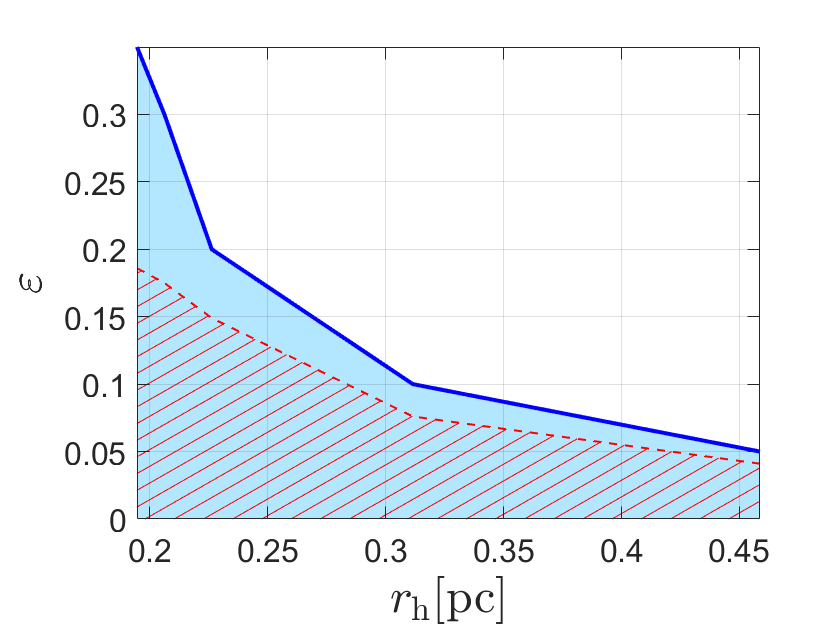}\end{minipage} & \begin{minipage}[c]{0.28\textwidth}\centering\includegraphics[width=0.95\textwidth]{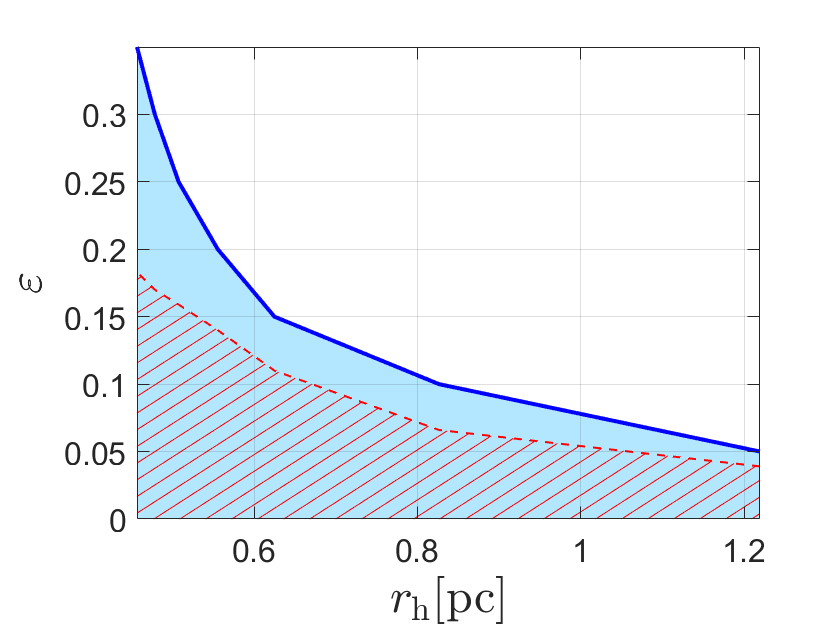}\end{minipage} & \begin{minipage}[c]{0.28\textwidth}\centering\includegraphics[width=0.95\textwidth]{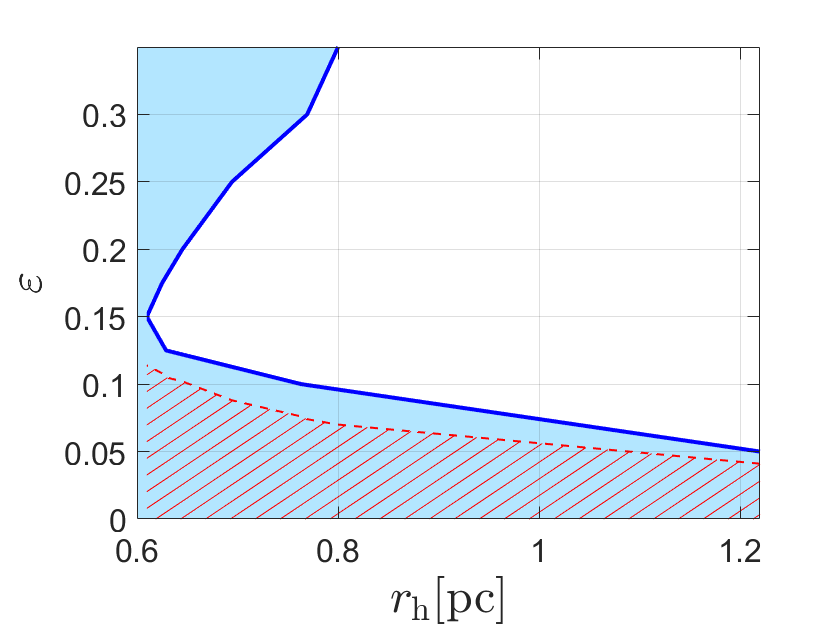}\end{minipage} \\
			$10^7{\rm M}_\odot$ & \begin{minipage}[c]{0.28\textwidth}\centering\includegraphics[width=0.95\textwidth]{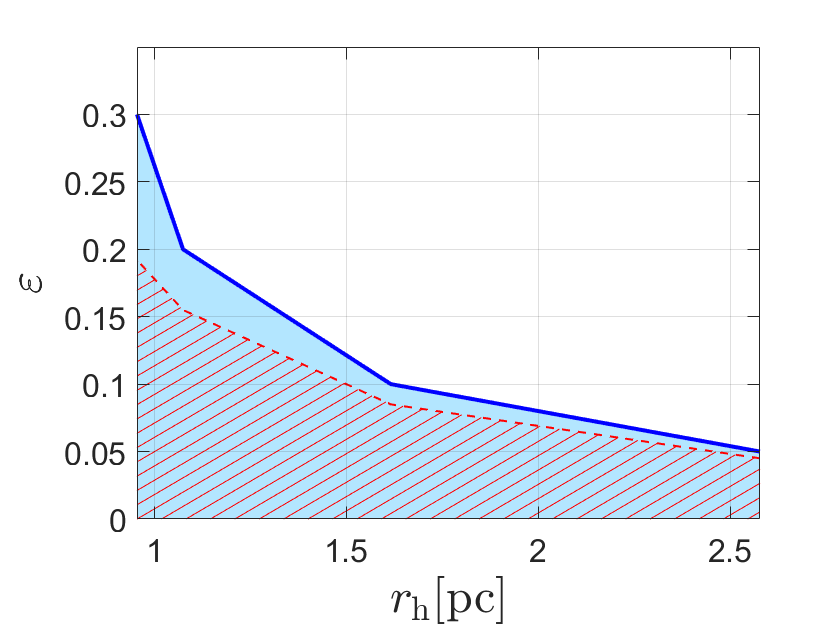}\end{minipage} & \begin{minipage}[c]{0.28\textwidth}\centering\includegraphics[width=0.95\textwidth]{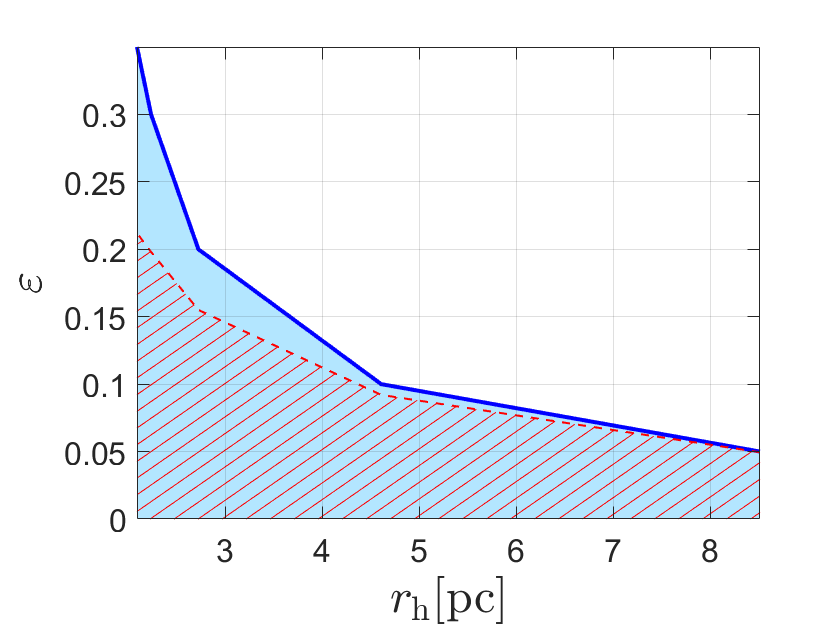}\end{minipage} & \begin{minipage}[c]{0.28\textwidth}\centering\includegraphics[width=0.95\textwidth]{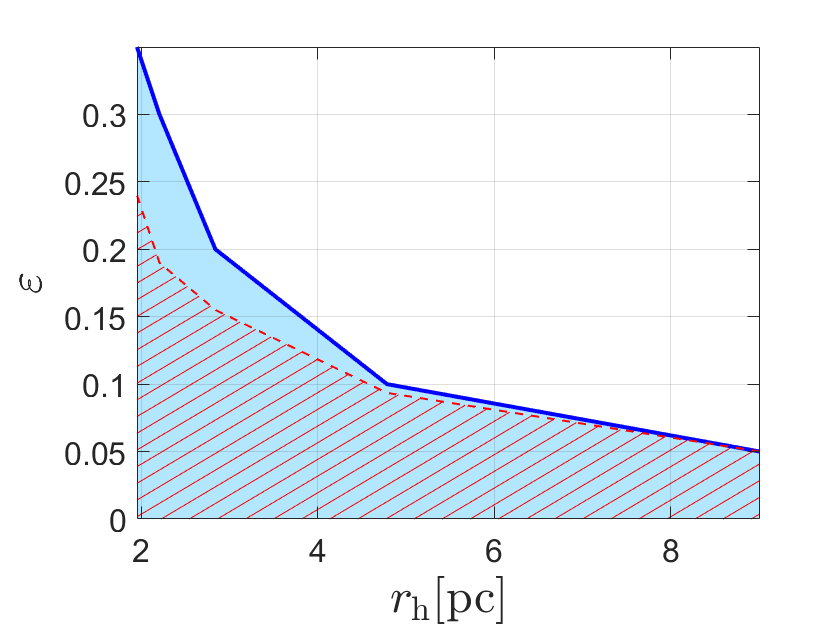}\end{minipage} \\
		\end{tabular}
		\caption{Range of parameters in which the proto-BHMF-shift operates. The rows correspond to different stellar-cluster total masses and the columns to different metallicities. The solid shaded blue region in the $\varepsilon$-$r_{\rm h}$ plane signifies their range in which the most massive BH grows by at least $10{\rm M}_\odot$. The diagonally hatched region signifies additionally the $\varepsilon$-$r_{\rm h}$ range for which an IMBH with a mass of $\sim 10^3{\rm M}_\odot$ can be generated. Notice that the subsolar and low-metallicity cases require an almost identical cluster compactness for the BHMF-shift to operate, but the solar metallicity case requires about double or higher compactness (x axis).}
		\label{fig:eps_rhf}
	\end{figure*}
	
	\section{Results}\label{sec:results}
	
	We implemented three different studies.
	Firstly, we performed ten runs of a proto-stellar cluster with an initial total mass of $10^6{\rm M}_\odot$ for three combinations of $\varepsilon$, $r_{\rm h}$ values for subsolar and low metallicity. These should represent the formation stage of  typical GCs with a final stellar mass of $\sim 10^5{\rm M}_\odot$. The results are summarized in Figures \ref{fig:BHmass_scatter_M1.0e6}, \ref{fig:BHMF_M1.0e6} and Table \ref{tab:dm_fit}. In the second study we provide a general picture of the effectiveness of our proto-BHMF-shift mechanism over a vast range of parameters. We ran multitude additional simulations to inspect the whole $\varepsilon$, $r_{\rm h}$ space regarding the possibility of a proto-BHMF-shift for vastly  different cluster initial total mass values of $10^{5-7}{\rm M}_\odot$, and for solar, subsolar, and low metallicities. These results are summarized in Figure \ref{fig:eps_rhf}.
	In the third study, we model a typical proto-stellar cluster of the recent JWST observations in the Cosmic Gems arc galaxy. These results are displayed in Figures \ref{fig:m_max_rh_Cosmic_Gems_M3e6} and \ref{fig:BHMF_cosmic_gems}. Finally we discuss the case of GW231123.
	
	\subsection{$M = 10^6{\rm M}_\odot$ proto-stellar clusters}
	
	In Figure \ref{fig:BHmass_scatter_M1.0e6} we depict scatter diagrams of the shifted BH masses with respect to the initial BH masses for the sets of values $\{r_{\rm h}, \varepsilon\} = \{0.1{\rm pc}, 0.06\}$, $\{0.07{\rm pc}, 0.1\}$ and $\{0.5{\rm pc}, 0.2\}$ at subsolar and low metallicity for a cluster with an initial total mass of $M = 10^6{\rm M}_\odot$. We varied $\varepsilon$ with $r_{\rm h}$ because the BHMF-shift requires lower $\varepsilon$ as $C$ is decreased ($r_{\rm h}$ is increased) to be effective, while it is also physically motivated: the star formation efficiency increases for more compact clusters (lower $r_{\rm h}$), where the gas is more condensed. These cluster mass and half-mass radius values correspond to compactness values of $C = 10$, $14.3$, $20$. The values $\{M,r_{\rm h}\} = \{10^6{\rm M}_\odot,0.07{\rm pc}\}$ correspond to Figures \ref{fig:BHMF_M1.0e6_C14.3_eps0.1_Zsub} and \ref{fig:BHMF_M1.0e6_C14.3_eps0.1_Zlow}, where it is evident that, for $\varepsilon = 0.1$, in the low-$Z$ case the whole upper BH mass gap is filled and spanned up to $150{\rm M}_\odot$. In the subsolar case, the BH masses reach up to $\sim 450{\rm M}_\odot$. These masses of $>150{\rm M}_\odot$ manifest as a tail in the shifted BHMF in Figure \ref{fig:BHMF_M1.0e6_C14.3_eps0.1_Zsub}. This is not special for the subsolar case, but can occur at any metallicity if the cluster is compact enough and/or the stellar formation efficiency is low enough, as we explain in detail below.

	We observe in all diagrams of Figure \ref{fig:BHmass_scatter_M1.0e6} that there are several BH branches with different behaviours. The main branch of CCSN-BH and direct-collapse-BH-low correspond to the lower progenitors' ZAMS branch of Figure \ref{fig:mass_BH_ZAMS} up to the first peak, which designates the upper BH mass cutoff and the direct collapse with the minimum mass loss. 
	We find that this branch follows a steep power law, described in Table \ref{tab:dm_fit}, with a power exponent in the range of $4.3 - 6.0$. This is the steeper mass increase, with the exception of the heavy mass tail of Figure \ref{fig:BHMF_M1.0e6_C14.3_eps0.1_Zsub}.
	The rest of the branches in Figure \ref{fig:BHmass_scatter_M1.0e6} correspond to the progenitors' ZAMS masses above the first peak of Figure \ref{fig:mass_BH_ZAMS} for which PPISN operates. 
	
	For intermediate cluster masses of $\sim 10^6{\rm M}_\odot$ the optimum metallicity for which the BHMF shift is mostly effective is somewhere in the subsolar regime. It should be low enough that the stellar winds are sufficiently weak but not so low as to invoke frequent PISN, which can drive violent gas expulsion. However, this is not necessarily the case for low ($\sim 10^5 {\rm M}_\odot$) and high ($\sim 10^7 {\rm M}_\odot$) cluster masses. A low-mass cluster is inefficient at frequently producing stars massive enough to generate PISN, so that low-metallicity clusters with weaker stellar winds can be more effective than subsolar-metallicity ones. At high cluster masses, on the other hand, PISNs are less effective in expelling the gas because of the intense gravity; therefore, low-metallicity clusters can be equally as effective as subsolar metallicity ones, or even more so, at generating a BHMF shift.
	
	Figure \ref{fig:BHMF_M1.0e6} depicts the BHMF shift for the same parameter values as in Figure \ref{fig:BHmass_scatter_M1.0e6}. 
	We show the fit of the shifted BHMF in three regimes: the whole mass range, the upper BH mass gap range, and the low-mass range below the mass gap. A fourth region of a heavy-mass ($m_{\rm BH} > 130 {\rm M}_\odot $) tail is optimally fit with a nearly constant curve in Figure \ref{fig:BHmass_scatter_M1.0e6_C14.3_eps0.1_Zsub}. The BHMF shift induces a steeper exponent with a change of $\Delta \alpha = 0.05-0.32$ with respect to the BH IMF, because BHs in the range of $\sim 25-55{\rm M}_\odot$ shift to higher masses. In all cases plotted in this figure, the upper BH mass gap is populated, at least partially. The plotted BHMFs for these parameters are indicative. Populating the upper BH mass gap and even the heavy tail can be reproduced for any case if we sufficiently decrease $r_{\rm h}$ or $\varepsilon$. The heavy tail can reach IMBH masses of $m_{\rm BH}\sim 10^3{\rm M}_\odot$. We quantify these thresholds in detail in the next subsection.
	
	\subsection{BHMF-shift thresholds}
	
	We investigated the possibility of BHMF shift over a vast range of possible cluster configurations. We ran multiple simulations for three cluster initial total mass values, $10^{5}{\rm M}_\odot$, $10^{6}{\rm M}_\odot$, and $10^{7}{\rm M}_\odot$, for solar, subsolar, and low metallicities, covering the whole $\varepsilon$, $r_{\rm h}$ range.
	In Figure \ref{fig:eps_rhf} we depict the range at which our proto-BHMF-shift mechanism operates in the $\varepsilon$-$r_{\rm h}$ plane. The solid filled region designates the conditions in which the initially heavier BH accretes at least $10{\rm M}_\odot$ within our simulation time (up to $99\%$ gas depletion). The diagonally hatched region signifies the range of values in which the formation of an IMBH, with a mass of $\sim 10^3{\rm M}_\odot$, is possible. 
	The case of $M = 10^6{\rm M}_\odot$ at low metallicity is of special interest. Remarkably, in this case the proto-BHMF shift operates for any $\varepsilon$ if $r_{\rm h} \leq 0.6$, and equivalently $C \geq 16.67$, well within the compactness range of observed proto-stellar clusters, $C_{\rm observed} \approx 10-50$. 
	This is explained by the form of the depletion timescale diagram \ref{fig:tau_low_PISN}. We observe there that the timescale stalls at precisely these compactness values, which are mostly relevant to $M = 10^6{\rm M}_\odot$. The reason for this stall is that energetic PISN explosions operate. Even higher compactness and/or lower $\varepsilon$ are not sufficient to increase the timescale within this compactness value range. However, the PISNs proceed fast and if the cluster reaches compactness values that allow it to survive the PISN bursts the timescale increases rapidly.
	Another interesting feature in this case (low-$Z$, mass $10^6{\rm M}_\odot$) is that even at higher $r_{\rm h} > 0.6{\rm pc}$ the proto-BHMF shift occurs for both low and high $\varepsilon$. For a fixed depletion timescale, as $\varepsilon$ increases the gas component gets more centrally condensed, allowing a similar accretion efficiency with lower $\varepsilon$, which implies more available gas for accretion. This effect does not occur in the rest of the cases because the depletion timescale does not stall.
	Our analysis shows that a significant BHMF shift is theoretically possible for all metallicities and for clusters with a vast mass range of $10^5-10^7{\rm M}_\odot$ as in Figure \ref{fig:eps_rhf}.

	\subsection{Cosmic Gems clusters}
	
	\begin{figure}[!tbp]
		\centering
		\includegraphics[width=0.8\columnwidth]{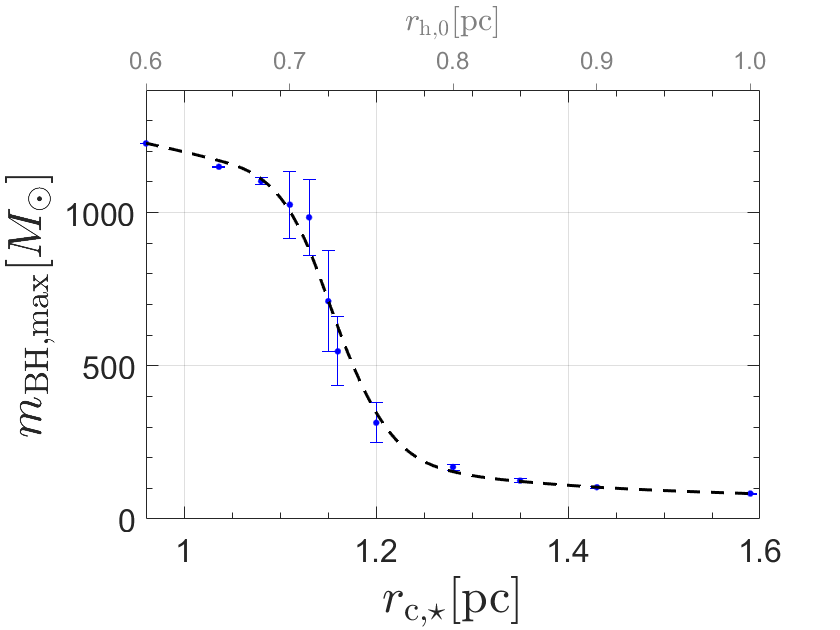}
		\caption{Maximum shifted individual BH mass of an initial $55{\rm M}_\odot$ BH in a stellar cluster with total (stellar) mass $M_\star = 10^6 {\rm M}_\odot$, for cluster sizes representative of the observed Cosmic Gems proto-stellar clusters. We assumed an initial gas-rich cluster with total $M = 3\cdot10^6 {\rm M}_\odot$ and star formation efficiency $\varepsilon = 0.35$ at low metallicity. The lower $x$ axis is the final (after gas depletion) Plummer radius, to be compared to the half-light radius of the observed stellar clusters, and the upper $x$ axis is the half-mass radius of the initial gas-rich cluster.}
		\label{fig:m_max_rh_Cosmic_Gems_M3e6}
	\end{figure}
	
	\begin{figure}[!htbp]
		\centering
		\begin{subfigure}[b]{0.85\columnwidth}
			\centering
			\includegraphics[width=0.9\textwidth]{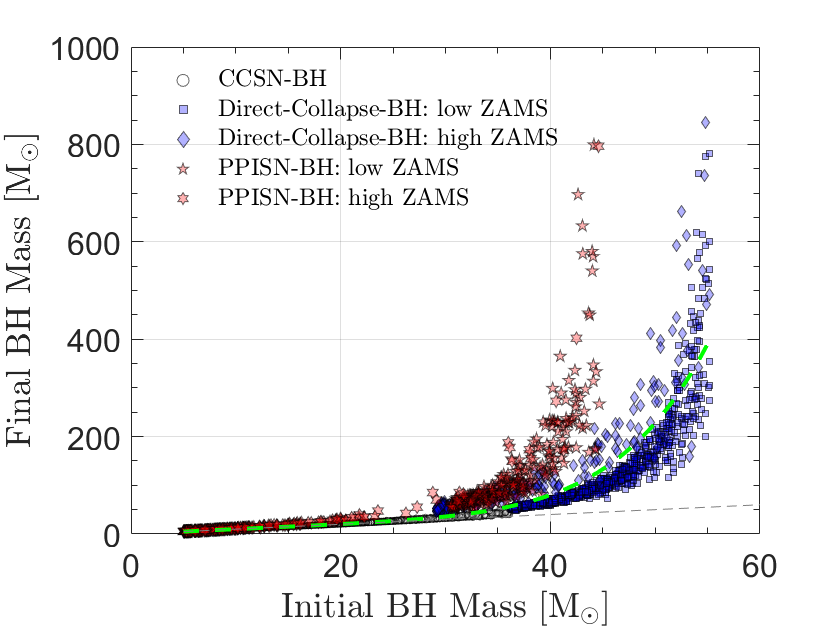}
			\caption{}
			\label{fig:BHmass_scatter_M3.0e6_C41.7_eps0.35_Zlow}
		\end{subfigure}
		\begin{subfigure}[b]{0.85\columnwidth}
			\centering
			\includegraphics[width=0.9\textwidth]{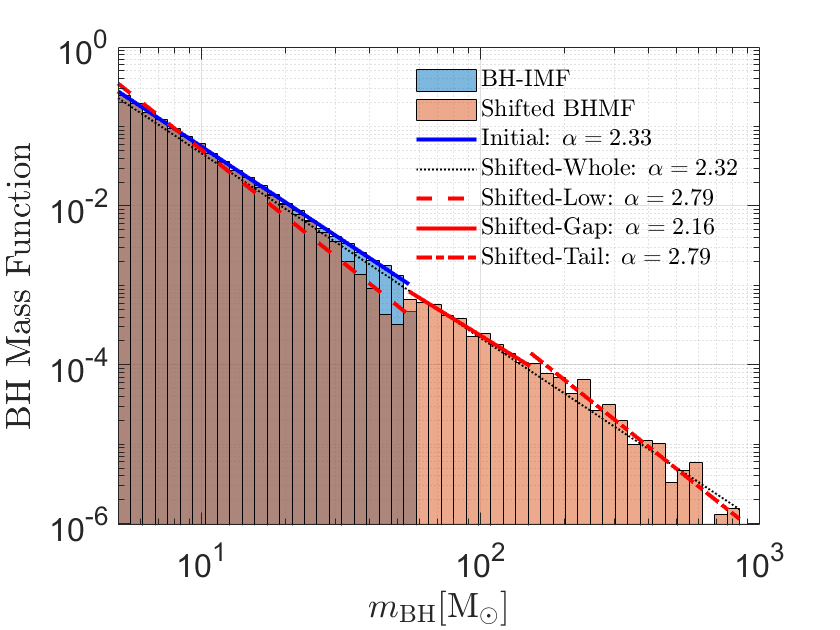}
			\caption{}
			\label{fig:BHMF_M3.0e6_C41.7_eps0.35_Zlow}
		\end{subfigure}
		\caption{Mass scatter diagram (a), and BHMF (b), for a cluster representing the observed Cosmic Gems proto-stellar clusters, following ten simulation runs. In particular, the initial parameters of the gas-rich cluster are $M = 3 \cdot 10^6 {\rm M}_\odot$, $r_{\rm h} = 0.72{\rm pc}$, $\varepsilon = 0.35$, $Z \simeq 0.01{\rm Z}_\odot$, and $\tau = 2.9 {\rm Myr}$, resulting in a stellar cluster after gas is depleted with $M_{\star} = 10^6{\rm M}_\odot$ and $r_{{\rm h},\star} = 1.15{\rm pc}$. The general trend of the individual BH mass increase is $\Delta m = (m_{\rm initial}/23 {\rm M}_\odot)^{6.7}$ depicted as the dashed green curve in (a). In (b), $\alpha$ denotes the exponent of each histogram fit.}
		\label{fig:BHMF_cosmic_gems}
	\end{figure}

	We modeled proto-stellar clusters, resembling the observed ones in the Cosmic Gems arc galaxy \citep{2024Natur.632..513A}. Their observed masses are in the range of $(1-4)\cdot 10^6{\rm M}_\odot$ and the half-light radii are $(0.7 - 1.7) {\rm pc}$.
	We adopted $\varepsilon = 0.35$, representing the upper end $\sim 0.3$ of observed star formation efficiencies in embedded clusters \citep{2003ARA&A..41...57L,2007MNRAS.380.1589B}, appropriate for the extreme densities and low metallicity in such massive, compact proto-stellar clusters as the Cosmic Gems ones.
	The observations do not allow one to conclusively determine the gas content of the clusters. Their observationally estimated age of $10-50 {\rm Myr}$ \citep{2024Natur.632..513A} suggests that they possibly are nearly at the final gas expulsion stage, with little gas remaining. However, the exact amount of gas remaining is uncertain while the galaxy itself is gas-rich. We assumed a low metallicity of $0.01 Z_\odot$, in accordance with the low metallicity expected for the Cosmic Gems arc galaxy \citep{2024Natur.632..513A}.
	
	We chose an initial total mass for the gas-rich cluster of $M = 3\cdot 10^6{\rm M}_\odot$, which means that $M_{\star} = 10^6{\rm M}\odot$ for the stellar component, which is at the low-mass end of the observations. 
	In our investigation, we cover a range of initial half-mass radii, $r_{\rm h} = 0.6-1{\rm pc}$, which result in stellar Plummer radii of $r_{{\rm c},\star} = 0.9-1.6{\rm pc}$ because of expansion due to mass loss. In Figure \ref{fig:m_max_rh_Cosmic_Gems_M3e6} we depict the maximum individual BH mass with respect to $r_{{\rm c},\star}$, to be compared with the observed half-light radius of Cosmic Gems clusters. For all these cluster sizes, the clusters have identical gas depletion timescales, due to the low-$Z$ PISN plateau (Figure \ref{fig:tau_low_PISN}),
	\begin{equation}
		\tau = 2.9 {\rm Myr},
	\end{equation}
	which implies that by $13{\rm Myr}$ $99\%$ of the gas has been expelled. This is the time at which we terminate the simulations. 
	We find that the upper BH mass gap is populated even for the most conservative cluster sizes, $\sim 1.3 - 1.6{\rm pc}$. 
	Remarkably, resulting stellar clusters with $r_{{\rm c},\star} \leq 1.15{\rm pc}$ can host IMBHs with masses of $\sim 10^3{\rm M}_\odot$.
	In Figure \ref{fig:BHMF_cosmic_gems} we present the results of ten runs for an initial half-mass radius of $r_{\rm h} = 0.72 {\rm pc}$ that gives a marginal size of $r_{{\rm c},\star} = 1.15{\rm pc}$, which is the maximum radius for which $\sim 10^3{\rm M}_\odot$ BHs can be formed. 
	This is a conservative value of the observed, effective half-light radii for a $\sim 10^6{\rm M}_\odot$ Cosmic Gems cluster (see Table 1 of \cite{2024Natur.632..513A}). 
	The BHMF shifts to higher masses (Figure \ref{fig:BHMF_M3.0e6_C41.7_eps0.35_Zlow}) populating the mass-gap with a negative power exponent of $\simeq 2.2$ fit within the mass-gap range. The power exponent of the tail, $m_{\rm BH}>130{\rm M}_\odot$, is $2.8$.
	There were generated, per cluster, $3082$ BHs (based on a stellar Kroupa IMF for progenitor stars), on average constituted of $2928$ CCSN-BHs, of which $1320$ escape due to recoil kick at birth, $102$ are direct-collapse-BHs, and $52$ are PPISN-BHs. The BHs accrete on average $4.8\cdot 10^4{\rm M}_\odot$ gas mass per cluster ($0.6\%$ of the initial gas reservoir).
	The maximum individual BH mass is $845{\rm M}_\odot$, and occurs for a direct-collapse-BH. 
	The general fit of the individual BH mass increase for all BHs, depicted as a dashed green line in Figure \ref{fig:BHmass_scatter_M3.0e6_C41.7_eps0.35_Zlow}, is
	$\Delta m_{\rm BH} = \left(m_{{\rm BH},{\rm birth}}/23{\rm M}_\odot \right)^{6.7}$.
	Excluding the outliers above mass $500{\rm M}_\odot$, the fit becomes $
	\Delta m_{\rm BH} = \left(m_{{\rm BH},{\rm birth}}/20.9{\rm M}_\odot \right)^{5.8}
	$. The distinct trends give a power exponent of $7.7$ and scale of $26.1{\rm M}_\odot$ for all direct-collapse-BHs and, respectively, $6.4$, $18.5{\rm M}_\odot$ for the PPISN-BHs.

	\subsection{GW231123}\label{sec:GW231123}
	
	Let us discuss the possibility that the BH components of the GW231123 \citep{2025arXiv250708219T} were generated with our proto-BHMF-shift mechanism in proto-stellar clusters. Their mass ranges, $103_{-52}^{+20} {\rm M}_\odot$ and $137_{-17}^{+22} {\rm M}_\odot$, are definitely within the typical proto-BHMF-shift range for a vast range of cluster parameters, including those of Cosmic Gems clusters. However, the critical data that could potentially discriminate between our accretion-driven mass growth in proto-stellar clusters and repeated mergers are the spins of the components BHs estimated as $a_* = 0.9_{-0.19}^{+0.10}$ and $a_* = 0.8_{-0.51}^{+0.20}$. Notice that these values are higher than the estimated equilibrium value of second-generation repeated mergers with an expected narrow peak at $a^{*}_{\rm merg} \sim 0.7$ \citep{2017ApJ...840L..24F} (higher-generation mergers in galactic nuclei could produce higher spins \citep{2019PhRvL.123r1101Y}). We argue here that these unusually high spin central values of GW231123 are nevertheless consistent with BH spin-up due to accretion, especially in the conditions in which our proto-BHMF-shift operates. 
	
	\citet{2013ApJ...762...68D} have shown that BHs with masses of $\lesssim 10^7 {\rm M}_\odot$ can maintain high spins, $a_{\star} \sim 0.8-0.9$, but still below the Thorne limit, $0.998$ \citep{1974ApJ...191..507T}, through chaotic accretion. For this to occur three conditions should be satisfied: (i) multiple randomly oriented accretion episodes, (ii) sufficiently fast BH-disk alignment, and (iii) sufficient disk angular momentum.
	
	In the case of our proto-BHMF-shift mechanism, all three conditions are satisfied. First, since the cluster turbulence has a correlation length of $\ell_{\rm corr} \approx r_{\rm h} >R_{\rm B}$ \citep{1992JFM...245....1B,2000ApJ...535..887K}, the time for turbulence and BH displacement to disrupt angular momentum coherence is the time needed for the BH to perform a random walk within $r_{\rm h}$; that is, $\tau_{\rm disruption} \sim r_{\rm h} / \sigma_{\rm BH} = \sqrt{m_{\rm BH}/\left\langle m_\star \right\rangle} \cdot r_{\rm h} / \sigma_\star \approx 0.4	{\rm Myr}\cdot (m_{\rm BH}/ 100{\rm M}_\odot)^{1/2}$. 
	The accretion timescale is $\tau_{\rm acc}\sim 1-10{\rm Myr}$ depending on local conditions. Multiple randomly oriented accretion events on the order of $\sim 20$ should occur for a $100{\rm M}_\odot$ BH. Second, the disk alignment timescale is $\tau_{\rm align} \approx 4\cdot 10^{-3}{\rm Myr}\cdot (m_{\rm BH}/100{\rm M}_\odot)^{-2/35}$ \citep{2013ApJ...762...68D}. This is negligible compared to $\tau_{\rm acc}$ and also yields $\tau_{\rm align}/\tau_{\rm disruption} \sim 10^{-2}$, allowing for disk alignment before disruption. 
	Finally, since we estimated $\sim 20$ realignment episodes, in order for the gas to add significant amount of angular momentum to the BH, the disk angular momentum, $J_{\rm disk} \approx M_{\rm disk} \sqrt{Gm_{\rm BH} r_{\rm out}}$, should be $J_{\rm disk} \gtrsim 0.05 J_{\rm BH}$, where  $J_{\rm BH} = a_{*} GM_{\rm BH}^2/c$. Assuming $M_{\rm disk} \sim \rho_{\rm gas} r_{\rm out}^3$ and in the cluster's core $\rho_{\rm gas} \simeq 10^6{\rm M}_\odot/{\rm pc}^3$, we get the condition for sufficient angular momentum transfer $r_{\rm out} / R_{\rm B} \gtrsim 10^{-2} \cdot a_\star^{2/7} \cdot (T/10^3{\rm K})\cdot (m_{\rm BH}/100{\rm M}_\odot)^{-4/7}$, which means that angular momentum transfer can occur inside the Bondi radius, $r_{\rm out} < R_{\rm B}$. Therefore, a sufficient amount of gas' angular momentum can be accumulated within the Bondi radius, especially for an initially low spin of $a_\star \sim 0.1$ and as gas cools down over time.
	
	Thus, we have enough evidence to indicate that the proto-BHMF-shift mechanism can generate a BH with spin at $a_{*} \sim 0.8-0.9$, consistent with the \citet{2013ApJ...762...68D} prediction for chaotic accretion onto (low-end) massive BHs. Nevertheless, we plan to investigate in detail the BH spin evolution of BHs with masses of $\sim 50-10^3{\rm M}_\odot$ within the turbulent environment of a gas-rich proto-stellar cluster in a subsequent work focused on this involved issue.
	
	\section{Conclusion}\label{sec:conclusion}
	
	We have calculated the conditions for a gas-rich proto-stellar cluster that can induce a significant BHMF shift, due to accretion. Our analysis includes: detailed calculations of gas depletion timescales incorporating stellar lifetimes, stellar winds, and supernova explosions; the generation of realistic BH populations with assigned progenitor masses; the time-dependent introduction of BHs into the simulation according to their formation times; and the modeling of stochastic perturbations from both gas and stellar components. These elements are crucial for determining the efficiency of BH mass growth over the brief ($\sim 10-20 \, {\rm Myr}$ in total) simulation times of interest.
	
	Our results identify the specific conditions in which a significant BHMF shift can occur, sufficient at least to populate the upper BH mass gap. At low metallicity with a cluster mass of $\sim 10^6 {\rm M}_\odot$, any star formation efficiency produces shifts if the cluster is sufficiently compact ($C > 16.7$). For other metallicity, mass, and compactness values, the star formation efficiency is constrained, as is shown in Figure \ref{fig:eps_rhf}. We find that heavy ($>150 {\rm M}_\odot$) BH tails (like in Figure \ref{fig:BHMF_M1.0e6_C14.3_eps0.1_Zsub}) and IMBH formation ($\sim10^3{\rm M}_\odot$) are possible for all compactness values if the star formation efficiency  is sufficiently low. Solar metallicity environments require significantly more compactness for the BHMF shift to be effective due to intense stellar winds. As a rule of thumb, the shift in individual BH masses follows a steep power law of
	$
	\Delta m_{\rm BH} \approx \left(m_{{\rm BH},{\rm birth}}/20{\rm M}_\odot \right)^{4-6}
	$,
	where the value of the power exponent depends on the compactness, metallicity, and star formation efficiency of the cluster.
	
	This analysis has direct observational implications. The observed Cosmic Gems proto-stellar clusters at high redshift likely experienced early accretion-driven BHMF shifts, generating $\gtrsim 25$ BHs with masses of $10^2 {\rm M}_\odot \lesssim m_{\rm BH} \lesssim 10^3 {\rm M}_\odot$ within their first $\lesssim 13 {\rm Myr}$. 
	
	Our mechanism can provide further an explanation for some of the heavy, mass-gap BHs ($>60{\rm M}_\odot$) detected by LIGO-Virgo-KAGRA.
	The key element that would allow us to distinguish BHs grown via repeated mergers from the ones grown via accretion is the spin measurement. Chaotic accretion in principle favors low spins. However, in the case in which an accretion disk can align with the BH spin quickly with respect to the timescale of erratic spin reorientation and is also massive enough to transfer significant angular momentum, \cite{2013ApJ...762...68D} have shown that an equilibrium is reached at $a_*\sim 0.8-0.9$ for massive BHs ($\lesssim 10^7{\rm M}_\odot$). This spin is close to the estimated central values of the recent GW231123 signal. 
	Our first estimates (see section \ref{sec:GW231123}) suggest that this may also occur for lower-mass BHs ($\sim 100{\rm M}_\odot$) in the physical conditions of a gas-rich proto-stellar cluster. However, this requires further investigation with more precise analytical calculations or simulations of BH spin, not performed here.
	
	This work opens several important avenues for future investigation. The “proto-BHMF-shift” mechanism provides a pathway for seeding high-redshift SMBH formation. The IMBHs produced through this channel represent promising targets for GW detection by the future LISA mission. Finally, we plan to investigate the distinctive spin signatures that our mechanism produces in accretion-driven BHs and to develop specific observational predictions to distinguish them from merger-formed BHs in current and future LIGO-Virgo-KAGRA GW data.
	
	\begin{acknowledgements}
		ZR is supported by the European Union's Horizon Europe Research and Innovation Programme under the Marie Sk\l{}odowska-Curie grant agreement No.~101149270--ProtoBH.
	\end{acknowledgements}
	
	\bibliography{BHMF_proto}
	\bibliographystyle{aa}	
	
\end{document}